\documentclass[pre,11pt,showpacs,superscriptaddress,aps]{revtex4}
\usepackage{amsfonts}
\usepackage[dvips]{graphicx}
\usepackage{epsfig}
\usepackage{fancyhdr}
\usepackage{color}
\usepackage{amsmath}
\usepackage{amssymb}
\begin{document}
\newtheorem{theorem}{Theorem}
\newtheorem{definition}{Definition}
\newtheorem{lemma}{Lemma}
\newtheorem{proposition}{Proposition}
\newtheorem{remark}{Remark}
\newtheorem{corollary}{Corollary}
\newtheorem{example}{Example}


\newcommand{\mbf}[1]{\mbox{\boldmath $#1$}}

\def\nequiv{\;{/}{\!\!\!\! }{\equiv}\;}
\def\R{\mbox{$I\!\!R$}}
\def\Z{\mbox{$I\!\!Z$}}
\def\b{\mbox{\mbf b}}
\def\c{\mbox{\mbf c}}
\def\0{\mbox{\mbf 0}}
\def\G{\mbox{\mbf G}}

\def\f{\mbox{\mbf f}}
\def\g{\mbox{\mbf g}}
\def\h{\mbox{\mbf h}}
\def\u{\mbox{\mbf u}}
\def\x{\mbox{\mbf x}}
\def\y{\mbox{\mbf y}}
\def\z{\mbox{\mbf z}}
\def\H{\mbox{\bf H}}

\def\F{\mbox{\bf F}}
\def\s{\mbox{\bf s}}
\def\e{\mbox{\bf e}}
\def\r{\mbox{\bf r}}

\def\1{\mbox{\mbf 1}}

\def\i{\mbox{\mbf i}}
\def\p{\mbox{\mbf p}}
\def\q{\mbox{\mbf q}}

\def\A{\mbox{\mbf A}}
\def\B{\mbox{\mbf B}}
\def\G{\mbox{\mbf G}}
\def\D{\mbox{\mbf D}}
\def\H{\mbox{\mbf H}}
\def\I{\mbox{\mbf I}}
\def\J{\mbox{\mbf J}}
\def\M{\mbox{\mbf M}}

\def\X{\mbox{\mbf X}}

\def\P{\mbox{\mbf P}}

\def\L{\mbox{\mbf L}}
\def\N{\mbox{\mbf N}}

\def\BibTeX{{\rm B\kern-.05em{\sc i\kern-.025em b}\kern-.08em
    T\kern-.1667em\lower.7ex\hbox{E}\kern-.125emX}}


\def\file#1{\texttt{#1}}

\title{Fractional Helmholtz and fractional wave equations with Riesz-Feller and generalized
Riemann-Liouville fractional derivatives}

\author{Ram K. Saxena}
\email{ram.saxena@yahoo.com} \affiliation{Department of Mathematics
and Statistics, Jai Narain Vyas University, Jodhpur - 342004, India}
\author{\v{Z}ivorad Tomovski}
\email{tomovski@pmf.ukim.mk} \affiliation{Faculty of Natural
Sciences and Mathematics, Institute of Mathematics, Saints Cyril and
Methodius University, 1000 Skopje, Macedonia}
\author{Trifce Sandev}
\email{trifce.sandev@drs.gov.de} \affiliation{Radiation Safety
Directorate, Partizanski odredi 143, P.O. Box 22, 1020 Skopje,
Macedonia}

\begin{abstract}
{\bf Abstract.} The objective of this paper is to derive
analytical solutions of fractional order Laplace, Poisson and Helmholtz equations in two variables derived from the
corresponding standard equations in two dimensions by replacing the integer order partial derivatives with fractional Riesz-Feller derivative and generalized Riemann-Liouville fractional derivative recently defined by Hilfer. The Fourier-Laplace transform method is employed to obtain the solutions in terms of Mittag-Leffler functions, Fox $H$-function and an integral operator containing a Mittag-Leffler function in the kernel. Results for fractional wave equation are presented as well. Some interesting special cases of these equations are considered. Asymptotic behavior and series representation of solutions are analyzed in detail. Many previously obtained results can be derived as special cases of those presented in this paper.


{\bf AMS 2000:} 26A33, 33E12, 33C60, 76R50, 44A10, 42A38.
\end{abstract}


\maketitle

\section{Introduction}

Fractional differential equations have been used in different fields
of science. To mention a few examples: fractional relaxation
equations have applications in the non-exponential relaxation theory
\cite{1,2,mainardi jcam,hilfer,hilfer fractals}; fractional
diffusion \cite{metzler1,metzler2} and fractional Fokker-Planck
equations \cite{metzler ffpe}, as well as fractional master
equations \cite{saxena master eq,hilfer csf,jumarie,pagnini}, in the
description of anomalous diffusive processes; fractional wave
equations has been used in the theory of vibrations of smart
materials in media where the memory effects can not be neglected
\cite{liang1,liang2}; etc.

In the present paper, we introduce a new generalization of
time-independent diffusion/wave equations, i.e. fractional Laplace,
fractional Poisson and fractional Helmholtz equations in two
variables in which both space variables $x$ and $y$ are of
fractional orders. We use fractional Riesz-Feller space
derivative \cite{feller} for the first variable, and generalized
Riemann-Liouville (R-L) fractional derivative
\cite{hilfer,hilfer_LT,tomovski na} for the second variable. Space-time
fractional wave equation with Riesz-Feller space derivative and generalized
R-L fractional time derivative is considered as well.
Similar time-dependent models are discussed earlier by many
authors, such as Haubold et al. \cite{haubold et al}, Saxena
\cite{saxena master eq}, Saxena et al.
\cite{saxena1,saxena,saxena2}, Tomovski et al. \cite{tomovski
na,tomovski cma,tomovski amc,tomovski and sandev nd}, etc.

Such generalized R-L time fractional derivative (or so-called
Hilfer-composite fractional time derivative in
\cite{mainardi-gorenflo,sandev jpa2011,tomovski et al,tomovski itsf,furati,tatar})
was used by Hilfer \cite{hilfer,hilfer2}, Sandev et al. \cite{sandev
jpa2011} and Tomovski et al. \cite{tomovski et al} in the analysis
of fractional diffusion equations, obtaining that such models may be
used in context of glass relaxation and aquifer problems.
Hilfer-composite time fractional derivative was also used by Saxena
et al. \cite{saxena arxiv} and Garg et al. \cite{garg} in the theory
of fractional reaction-diffusion equations, where the obtained
results are presented through Mittag-Leffler (M-L) and Fox
$H$-functions. Furthermore, an operational method for solving
differential equations with the Hilfer-composite fractional
derivative is presented in \cite{hilfer_LT,kim fcaa,kim2}. From the
other side, Riesz-Feller fractional derivative has been used in
analysis of space-time fractional diffusion equations by Mainardi,
Pagnini and Saxena \cite{mainardi pagnini saxena} and Tomovski et
al. \cite{tomovski et al}, where they expressed the solutions in terms of
Fox $H$-function. It is shown that space fractional diffusion
equation with fractional Riesz-Feller space derivative \cite{compte}
gives same results as those obtained from the continuous time random
walk theory for L\'{e}vy flights \cite{metzler1,metzler2}. A
numerical scheme for solving fractional diffusion equation with
Hilfer-composite fractional time derivative and Riesz-Feller space
fractional derivative is elaborated in \cite{tomovski et al}. Furthermore,
the quantum fractional Riesz-Feller derivative has been used by
Luchko et al. \cite{luchko jmp,luchko epj} in the Schr\"{o}dinger equation
for a free particle and a particle in an infinite potential well. Local fractional
derivative operators have been used as well \cite{hmsrivastava1} in Helmholtz
and diffusion equations.

The paper is organized as following. In Section II we give an introduction to
the fractional derivatives and integrals used in the paper. Fractional form of
the Laplace and Poisson equations in two variables are considered in Section III.
We give analytical results for different forms of the boundary conditions and for
the source term. Asymptotic behavior and series representation of solutions are
given. We also give remarks on the general space-time fractional wave equation for
a vibrating string with fractional Riesz-Feller space derivative and
Hilfer-composite fractional time derivative. In Section IV we analyze the fractional
Helmholtz equation for different forms of the boundary conditions and source term.
The obtained results are of general character and include those recently given by
Thomas \cite{thomas ijde}. Conclusions are given in Section V. At the end of the paper
in an Appendix we give definitions, relations, and some properties of M-L functions
and Fox $H$-function.

\section{Fractional derivatives and integrals}

The Riesz-Feller fractional derivative of order $\alpha$ and
skewness $\theta$ is defined by the following Fourier transform
formula \cite{feller}
\begin{equation}\label{feller}
\mathcal{F}\left[{_{x}}D_{\theta}^{\alpha}f(x)\right](\kappa)=-\psi_{\alpha}^{\theta}(\kappa)\mathcal{F}\left[f(x)\right](\kappa),
\end{equation}
where
\begin{equation}\label{fourier transform}
\mathcal{F}\left[f(x)\right](\kappa)=\hat{f}(\kappa)=\int_{-\infty}^{\infty}f(x)e^{\imath\kappa
x}\mathrm{d}x, \quad
\mathcal{F}^{-1}\left[\hat{f}(\kappa)\right](x)=\frac{1}{2\pi}\int_{-\infty}^{\infty}\hat{f}(\kappa)e^{-\imath\kappa
x}\mathrm{d}\kappa,
\end{equation}
are Fourier transform and inverse Fourier transform, respectively, and $\psi_{\alpha}^{\theta}(\kappa)$ is given by
\begin{equation}\label{psi}
\psi_{\alpha}^{\theta}(\kappa)=|\kappa|^{\alpha}\exp\left[\imath\mathrm{sign}(\kappa)\frac{\theta\pi}{2}\right],
\quad 0<\alpha\leq2, \quad |\theta|\leq\min\{\alpha,2-\alpha\}.
\end{equation}
Riesz-Feller fractional derivative is a pseudo-differential operator
whose symbol $-\psi_{\alpha}^{\theta}(\kappa)$ is the logarithm of
the characteristic function of a general L\'{e}vy strictly stable
probability density with stability index $\alpha$ and asymmetry
parameter $\theta$ (for details, see Mainardi, Pagnini and Saxena
\cite{mainardi pagnini saxena}). For $\theta=0$ one obtains Riesz
fractional derivative
${_{x}}D_{0}^{\alpha}=-\left(-\frac{\mathrm{d}^{2}}{\mathrm{d}x^{2}}\right)^{\alpha/2}$,
for which
\begin{equation}\label{feller_theta0}
\mathcal{F}\left[{_{x}}D_{0}^{\alpha}f(x)\right](\kappa)=-|\kappa|^{\alpha}\mathcal{F}\left[f(x)\right](\kappa).
\end{equation}
This special case has been used in the theory of L\'{e}vy flights
\cite{metzler1,metzler2}.

In this paper we also use the quantum fractional Riesz-Feller derivative
${_{x}}D_{\theta}^{*,\alpha}$ of order $\alpha$ and skewness
$\theta$, which is defined as a pseudo-differential operator with a
symbol $\psi_{\alpha}^{\theta}(\kappa)$ given by \cite{luchko
jmp,luchko epj}
\begin{equation}\label{quantum RF}
\mathcal{F}\left[{_{x}}D_{\theta}^{*,\alpha}f(x)\right](\kappa)=\psi_{\alpha}^{\theta}(\kappa)\mathcal{F}\left[f(x)\right](\kappa).
\end{equation}
Note that the quantum fractional Riesz-Feller derivative is the
Riesz-Feller fractional derivative (\ref{feller}) multiplied by
$-1$. Thus, the obtained solutions which correspond to the case of fractional Riesz-Feller space derivative (\ref{quantum RF})
can be easily transformed to those obtained in a case where the quantum fractional Riesz-Feller space derivative (\ref{feller}) is applied.

The R-L fractional integral is defined by \cite{samko,hilfer,35}
\begin{equation}\label{Eq.7}
\left({_{y}}I_{a+}^{\mu}f\right)(y)=\frac{1}{\Gamma(\mu)}\int_{a}^y
\frac{f(y')}{(y-y')^{1-\mu}}\textrm{d}y', \quad y>a, \quad
\Re(\mu)>0.
\end{equation}
For $\mu=0$, this is the identity operator,
$\left({_{y}}I_{a+}^{0}f\right)(y)=f(y)$. Similarly, R-L fractional
derivative is defined by \cite{samko,hilfer,35}
\begin{equation}\label{Eq.7_1}
\left({_{y}}D_{a+}^{\mu}f\right)(y)=\left(\frac{\mathrm{d}}{\mathrm{d}y}\right)^n\left({_{y}}I_{a+}^{n-\mu}f\right)(y),
\quad \Re(\mu)>0, \quad n=\left[\Re(\mu)\right]+1,
\end{equation}
where $[\Re(\mu)]$ denotes the integer part of the real number
$\Re(\mu)$. Hilfer generalized the fractional derivative
(\ref{Eq.7_1}) by the following fractional derivative of order
$0<\mu\leq1$ and type $0\leq\nu\leq1$ \cite{hilfer}:
\begin{equation}\label{hilfer}
\left({_{y}}D_{a+}^{\mu,\nu}f\right)(y)=\left({_{y}}I_{a+}^{\nu(1-\mu)}\frac{\mathrm{d}}{\mathrm{d}y}\left({_{y}}I_{a+}^{(1-\nu)(1-\mu)}f\right)\right)(y).
\end{equation}
Note that when $0<\mu\leq1$, $\nu=0$, $a=0$, the generalized R-L
fractional derivative (\ref{hilfer}) would correspond to the
classical R-L fractional derivative \cite{35,samko,hilfer}
\begin{equation}\label{R-L derivative0}
\left({{^{RL}}{_{y}}}D_{0+}^{\mu}f\right)(y)=\frac{\mathrm{d}}{\mathrm{d}y}\left({_{y}}I_{0+}^{(1-\mu)}f\right)(y).
\end{equation}
Conversely, when $0<\mu\leq1$, $\nu=1$, $a=0$, it corresponds to the
Caputo fractional derivative \cite{caputo}
\begin{equation}\label{Caputo derivative}
\left({^{C}}{_{y}}D_{0+}^{\mu}f\right)(y)=\left({_{y}}I_{0+}^{(1-\mu)}\frac{\mathrm{d}}{\mathrm{d}y}f\right)(y).
\end{equation}

The difference between fractional derivatives of different types
becomes apparent when we consider their Laplace transform. In
Ref.~\cite{hilfer} it is found for $0<\mu <1$ that
\begin{equation}\label{hilfer_laplace}
\mathcal{L}\left[{_{y}}D_{0+}^{\mu,\nu}f(y)\right](s)=s^\mu\mathcal{L}\left[f(y)\right](s)-s^{\nu(\mu-1)}\left({_{y}}I_{0+}^{(1-\nu)(1-\mu)}f\right)(0+),
\end{equation}
where the initial-value term
$\left({_{y}}I_{0+}^{(1-\nu)(1-\mu)}f\right)(0+)$ is evaluated in
the limit $y\rightarrow0+$, in the space of summable Lebesgue
integrable functions
\begin{equation}\label{lebesque}
L(0,\infty)=\left\{f:\|f\|_{1}=\int_{0}^{\infty}|f(y)|\mathrm{d}y<\infty\right\}.
\end{equation}

Hilfer, Luchko and Tomovski generalized Hilfer-composite derivative
(\ref{hilfer}) to order $n-1<\mu\leq n$ ($n\in N^{+}$) and type $0\leq\nu\leq1$ in
the following way \cite{hilfer_LT}:
\begin{equation}\label{hilfer l t}
\left({_{y}}D_{a+}^{\mu,\nu}f\right)(y)=\left({_{y}}I_{a+}^{\nu(n-\mu)}\frac{\mathrm{d}^{n}}{\mathrm{d}y^{n}}\left({_{y}}I_{a+}^{(1-\nu)(n-\mu)}f\right)\right)(y).
\end{equation}
Its Laplace transform is recently given by Tomovski \cite{tomovski
na}
\begin{equation}\label{generalized hilfer_laplace}
\mathcal{L}\left[{_{y}}D_{0+}^{\mu,\nu}f(y)\right](s)=s^\mu\mathcal{L}\left[f(y)\right](s)
-\sum_{k=0}^{n-1}s^{n-k-\nu(n-\mu)-1}\left[\frac{\mathrm{d}^{k}}{\mathrm{d}y^{k}}\left({_{y}}I_{0+}^{(1-\nu)(n-\mu)}f\right)\right](0+),
\end{equation}
where initial-value terms
$\left[\frac{\mathrm{d}^{k}}{\mathrm{d}y^{k}}\left({_{y}}I_{0+}^{(1-\nu)(n-\mu)}f\right)\right](0+)$
are evaluated in the limit $y\rightarrow0+$.

Various operators for fractional integration were investigated by
Srivastava and Saxena \cite{srivastava2}. Srivastava and Tomovski
\cite{34} introduced an integral operator
$(\mathcal{E}_{a+;\alpha,\beta}^{\omega;\gamma,\kappa}\varphi)(y)$
of form
\begin{equation}\label{Eq.11}
({_{y}}\mathcal{E}_{a+;\alpha,\beta}^{\omega;\gamma,\kappa}\varphi)(y)=\int_{a}^y
(y-\xi)^{\beta-1}E_{\alpha,\beta}^{\gamma,\kappa}(\omega
(y-\xi)^\alpha)\varphi(\xi)\mathrm{d}\xi,
\end{equation}
where $E_{\alpha,\beta}^{\gamma,\kappa}(z)$ is the four parameter
M-L function (\ref{Eq.12}). In case when $\omega=0$ the integral
operator (\ref{Eq.11}) would correspond to the classical R-L
integral operator. For $\kappa=1$ integral operator (\ref{Eq.11})
becomes the Prabhakar integral operator
$\left({_{y}}\mathcal{E}_{a+;\alpha,\beta}^{\omega;\gamma}\varphi\right)(y)$
\cite{prabhakar}, which was extensively investigated by Kilbas, Saigo and Saxena \cite{kilbas}, and will be used here with $\gamma=1$ for
representation of solutions. These generalized integral operators
was shown to appear in the expression of solutions of fractional
diffusion/wave equations with source terms \cite{sandev
jpa2010,sandev jpa2011,tomovski cma,tomovski amc,tomovski and sandev
nd}.

\section{Fractional Laplace and fractional Poisson equations}

In this section we investigate generalized form of the Laplace
equation for the field variable $N(x,y)$ in two dimensions
\begin{equation}\label{Laplace eq}
\frac{\partial^{2}}{\partial
x^{2}}N(x,y)+\frac{\partial^{2}}{\partial y^{2}}N(x,y)=0,
\end{equation}
on the upper half plane $y\geq0$ and $-\infty<x<\infty$, with
boundary conditions
\begin{subequations}
\begin{equation}\label{boundary1 Laplace}
N(x,0+)=f(x), \quad \frac{\mathrm{d}}{\mathrm{d}y}N(x,0+)=g(x),
\end{equation}
\begin{equation}\label{boundary2 Laplace}
\lim_{x\rightarrow\pm\infty}N(x,y)=0.
\end{equation}
\end{subequations}
Since there is no dependence on time variable, Laplace equation
gives the steady-state solution of, for example, diffusion/heat
conduction and wave equations. Thus, initial conditions are not
required, only we use boundary conditions, which may be defined in a
different ways. Therefore, Laplace equation in two dimensions
(\ref{Laplace eq}) may arise in analysis of two dimensional
steady-state diffusion/heat conduction, static deflection of a
membrane, electrostatic potential, etc.

If in the Laplace equation in two dimensions (\ref{Laplace eq}) we
add a source term $\Phi(x,y)$, then it becomes Poisson equation
\begin{equation}\label{Poisson eq}
\frac{\partial^{2}}{\partial
x^{2}}N(x,y)+\frac{\partial^{2}}{\partial y^{2}}N(x,y)=\Phi(x,y).
\end{equation}
This equation has applications in different field of science, such
as gravitation theory, electromagnetism, elasticity, etc. For
example, $N(x,y)$ may be interpreted as a temperature field variable
subject to external force (source) $\Phi(x,y)$.

Before to formulate the corresponding fractional form of the Laplace
equation (\ref{Laplace eq}) and Poisson equation (\ref{Poisson eq})
we prove the following Lemmas.

\begin{lemma}\label{lemma_ML}
Let $1<\mu\leq2$, $0\leq\nu\leq1$, $\varsigma\geq0$ and $\hat{r}(\kappa)$ is a
given function. Then the following relation holds true
\begin{equation}\label{lemma1}
\mathcal{L}^{-1}\left[\frac{s^{\varsigma-\nu(2-\mu)}}{s^\mu\pm\hat{r}(\kappa)}\right](y)=y^{1-(1-\nu)(2-\mu)-\varsigma}E_{\mu,2-(1-\nu)(2-\mu)-\varsigma}\left(\mp\hat{r}(\kappa)y^{\mu}\right),
\end{equation}
where $E_{\alpha,\beta}(z)$ is the two parameter M-L function
(\ref{Eq.5}).
\end{lemma}

\textit{Proof.} From relation (\ref{Eq.6}), we directly prove Lemma
\ref{lemma_ML}.

\begin{lemma}\label{lemma_ML convolution}
Let $1<\mu\leq2$ and $\hat{r}(\kappa)$ and $\hat{\Phi}(\kappa,y)$ are
given functions. Then the following relation holds true
\begin{equation}\label{lemma2}
\mathcal{L}^{-1}\left[\frac{1}{s^\mu\pm\hat{r}(\kappa)}\mathcal{L}\left[\hat{\Phi}(\kappa,y)\right](\kappa,s)\right](\kappa,y)=\left(\mathcal{E}_{0+;\mu,\mu}^{\mp\hat{r}(\kappa);1}\hat{\Phi}\right)(\kappa,y),
\end{equation}
where $\mathcal{E}_{0+;\mu,\mu}^{\mp\hat{r}(\kappa);1}\tilde{\Phi}$ is
the Prabhakar integral operator (see definition (\ref{Eq.11})) and
$\hat{\Phi}(\kappa,y)$ is a given function.
\end{lemma}

\textit{Proof.} From relation (\ref{Eq.6}) it follows that
\begin{equation}\label{lemma2_1}
\frac{1}{s^\mu\pm\hat{r}(\kappa)}=\mathcal{L}\left[y^{\mu-1}E_{\mu,\mu}\left(\mp\hat{r}(\kappa)y^{\mu}\right)\right](\kappa,s).
\end{equation}
Thus by applying the convolution theorem of the Laplace transform
one obtains
\begin{equation}\label{lemma2_2}
\mathcal{L}^{-1}\left[\frac{1}{s^\mu\pm\hat{r}(\kappa)}\mathcal{L}\left[\hat{\Phi}(\kappa,t)\right](\kappa,s)\right](\kappa,t)=\int_0^y(y-\xi)^{\mu-1}E_{\mu,\mu}^{1}\left(\mp\hat{r}(\kappa)(y-\xi)^\mu\right)\hat{\Phi}(\kappa,\xi)\mathrm{d}\xi,
\end{equation}
from where we obtain the proof of Lemma \ref{lemma_ML
convolution}.

\begin{theorem}\label{FPeq}
The solution of the following fractional Poisson equation
\begin{equation}\label{th_FPeq}
{_{x}}D_{\theta}^{\alpha}N(x,y)+{_{y}}D_{0+}^{\mu,\nu}N(x,y)=\Phi(x,y),
\end{equation}
where $x\in R$, $y\in R^{+}$, $1<\alpha\leq2$,
$|\theta|\leq\min\{\alpha,2-\alpha\}$, $1<\mu\leq2$,
$0\leq\nu\leq1$, with boundary conditions
\begin{subequations}
\begin{equation}\label{boundary1 FPeq}
\left({_{y}}I_{0+}^{(1-\nu)(2-\mu)}N\right)(x,0+)=f(x), \quad
\left(\frac{\mathrm{d}}{\mathrm{d}y}\left({_{y}}I_{0+}^{(1-\nu)(2-\mu)}N\right)\right)(x,0+)=g(x),
\end{equation}
\begin{equation}\label{boundary2 FPeq}
\lim_{x\rightarrow\pm\infty}N(x,y)=0,
\end{equation}
\end{subequations}
is given by
\begin{eqnarray}\label{th_FPeq_sol}
N(x,y)&&=\frac{y^{-(1-\nu)(2-\mu)}}{2\pi}\int_{-\infty}^{\infty}E_{\mu,1-(1-\nu)(2-\mu)}\left(y^{\mu}\psi_{\alpha}^{\theta}(\kappa)\right)\hat{f}(\kappa)e^{-\imath\kappa
x}\mathrm{d}\kappa\nonumber\\&&
+\frac{y^{1-(1-\nu)(2-\mu)}}{2\pi}\int_{-\infty}^{\infty}E_{\mu,2-(1-\nu)(2-\mu)}\left(y^{\mu}\psi_{\alpha}^{\theta}(\kappa)\right)\hat{g}(\kappa)e^{-\imath\kappa
x}\mathrm{d}\kappa\nonumber\\&&
+\frac{1}{2\pi}\int_{-\infty}^{\infty}\int_{0}^{y}
(y-\xi)^{\mu-1}E_{\mu,\mu}\left((y-\xi)^{\mu}\psi_{\alpha}^{\theta}(\kappa)\right)\hat{\Phi}(\kappa,\xi)e^{-\imath\kappa
x}\mathrm{d}\xi\mathrm{d}\kappa\nonumber\\
&&=\frac{y^{-(1-\nu)(2-\mu)}}{2\pi}\int_{-\infty}^{\infty}E_{\mu,1-(1-\nu)(2-\mu)}\left(y^{\mu}\psi_{\alpha}^{\theta}(\kappa)\right)\hat{f}(\kappa)e^{-\imath\kappa
x}\mathrm{d}\kappa\nonumber\\&&
+\frac{y^{1-(1-\nu)(2-\mu)}}{2\pi}\int_{-\infty}^{\infty}E_{\mu,2-(1-\nu)(2-\mu)}\left(y^{\mu}\psi_{\alpha}^{\theta}(\kappa)\right)\hat{g}(\kappa)e^{-\imath\kappa
x}\mathrm{d}\kappa\nonumber\\&&+\frac{1}{2\pi}\int_{-\infty}^{\infty}\left({_{y}}\mathcal{E}_{0+;\mu,\mu}^{\psi_{\alpha}^{\theta}(\kappa);1}\hat{\Phi}\right)(\kappa,y)
e^{-\imath\kappa x}\mathrm{d}\kappa,
\end{eqnarray}
where
$\hat{\Phi}(\kappa,y)=\mathcal{F}\left[\Phi(x,y)\right](\kappa,y)$.
\end{theorem}

\textit{Proof.} From the Laplace transform (\ref{generalized
hilfer_laplace}) to equation (\ref{th_FPeq}) and the boundary
conditions (\ref{boundary1 FPeq}), it follows
\begin{equation}\label{sol2}
{_{x}}D_{\alpha}^{\theta}\tilde{N}(x,s)+s^{\mu}\tilde{N}(x,s)-s^{1-\nu(2-\mu)}f(x)-s^{-\nu(2-\mu)}g(x)=\tilde{\Phi}(x,s),
\end{equation}
where $\tilde{N}(x,s)=\mathcal{L}\left[N(x,y)\right](x,s)$. By applying
Fourier transform (\ref{quantum RF}) to relation (\ref{sol2}) we find
\begin{equation}\label{sol22}
\hat{\tilde{N}}(\kappa,s)=\frac{s^{1-\nu(2-\mu)}}{s^{\mu}-\psi_{\alpha}^{\theta}(\kappa)}\hat{f}(\kappa)
+\frac{s^{-\nu(2-\mu)}}{s^{\mu}-\psi_{\alpha}^{\theta}(\kappa)}\hat{g}(\kappa)+\frac{1}{s^{\mu}-\psi_{\alpha}^{\theta}(\kappa)}\hat{\tilde{\Phi}}(\kappa,s),
\end{equation}
where
$\hat{\tilde{\Phi}}(\kappa,s)=\mathcal{F}\left[\tilde{\Phi}(x,s)\right](\kappa,s)$.
Employing the results from Lemma \ref{lemma_ML} and Lemma
\ref{lemma_ML convolution}, by inverse Fourier transform we obtain
solution (\ref{th_FPeq_sol}). Thus, we finish with the proof of Theorem \ref{FPeq}.

\begin{remark}\label{FPeq_quantum}
If in equation (\ref{th_FPeq}) instead of fractional Riesz-Feller derivative we use quantum fractional Riesz-Feller derivative we obtain the following equation
\begin{equation}\label{th_FPeq_quantum}
{_{x}}D_{\theta}^{*,\alpha}N(x,y)+{_{y}}D_{0+}^{\mu,\nu}N(x,y)=\Phi(x,y).
\end{equation}
For same boundary conditions as those used in Theorem (\ref{FPeq}), we obtain the solution in the following form
\begin{eqnarray}\label{th_FPeq_sol_quantum}
N(x,y)&&=\frac{y^{-(1-\nu)(2-\mu)}}{2\pi}\int_{-\infty}^{\infty}E_{\mu,1-(1-\nu)(2-\mu)}\left(-y^{\mu}\psi_{\alpha}^{\theta}(\kappa)\right)\hat{f}(\kappa)e^{-\imath\kappa
x}\mathrm{d}\kappa\nonumber\\&&
+\frac{y^{1-(1-\nu)(2-\mu)}}{2\pi}\int_{-\infty}^{\infty}E_{\mu,2-(1-\nu)(2-\mu)}\left(-y^{\mu}\psi_{\alpha}^{\theta}(\kappa)\right)\hat{g}(\kappa)e^{-\imath\kappa
x}\mathrm{d}\kappa\nonumber\\&&+\frac{1}{2\pi}\int_{-\infty}^{\infty}\left({_{y}}\mathcal{E}_{0+;\mu,\mu}^{-\psi_{\alpha}^{\theta}(\kappa);1}\hat{\Phi}\right)(\kappa,y)
e^{-\imath\kappa x}\mathrm{d}\kappa.
\end{eqnarray}
\end{remark}

\begin{corollary}
If we consider source term of form
$\Phi(x,y)=\delta(x)\frac{y^{-\beta}}{\Gamma(1-\beta)}$, the
solutions of fractional equations (\ref{th_FPeq}) and (\ref{th_FPeq_quantum}) are given by
\begin{eqnarray}\label{th_FPeq_sol delta power}
N(x,y)&&=\frac{y^{-(1-\nu)(2-\mu)}}{2\pi}\int_{-\infty}^{\infty}E_{\mu,1-(1-\nu)(2-\mu)}\left(\pm y^{\mu}\psi_{\alpha}^{\theta}(\kappa)\right)\hat{f}(\kappa)e^{-\imath\kappa
x}\mathrm{d}\kappa\nonumber\\&&
+\frac{y^{1-(1-\nu)(2-\mu)}}{2\pi}\int_{-\infty}^{\infty}E_{\mu,2-(1-\nu)(2-\mu)}\left(\pm y^{\mu}\psi_{\alpha}^{\theta}(\kappa)\right)\hat{g}(\kappa)e^{-\imath\kappa
x}\mathrm{d}\kappa\nonumber\\&&
+\frac{y^{\mu-\beta}}{2\pi}\int_{-\infty}^{\infty}E_{\mu,\mu-\beta+1}\left(\pm y^{\mu}\psi_{\alpha}^{\theta}(\kappa)\right)e^{-\imath\kappa
x}\mathrm{d}\kappa\nonumber\\&&
=\frac{y^{-(1-\nu)(2-\mu)}}{2\pi}\int_{-\infty}^{\infty}E_{\mu,1-(1-\nu)(2-\mu)}\left(\pm y^{\mu}\psi_{\alpha}^{\theta}(\kappa)\right)\hat{f}(\kappa)e^{-\imath\kappa
x}\mathrm{d}\kappa\nonumber\\&&
+\frac{y^{1-(1-\nu)(2-\mu)}}{2\pi}\int_{-\infty}^{\infty}E_{\mu,2-(1-\nu)(2-\mu)}\left(\pm y^{\mu}\psi_{\alpha}^{\theta}(\kappa)\right)\hat{g}(\kappa)e^{-\imath\kappa
x}\mathrm{d}\kappa\nonumber\\&&
+\frac{y^{\mu-\beta}}{|x|}H_{3,3}^{2,1}\left[\mp\frac{|x|^{\alpha}}{y^{\mu}e^{\imath\frac{\theta\pi}{2}}}\left|\begin{array}{c
l}
     (1,1), (1+\mu-\beta,\mu), (1,\frac{\alpha}{2})\\
    (1,\alpha), (1,1), (1,\frac{\alpha}{2})
  \end{array}\right.\right],
\end{eqnarray}
where the upper signs in the solution correspond to the case of fractional Riesz-Feller derivative and lower signs to quantum fractional Riesz-Feller derivative.
\end{corollary}

\begin{example}
If we consider $\Phi(x,y)=\delta(x)\delta(y)$ and boundary
conditions $f(x)=\delta(x)$, $g(x)=0$, for $\theta=0$, from
relations (\ref{HML}) and (\ref{cosine H}), we obtain solutions
(\ref{th_FPeq_sol}) and (\ref{th_FPeq_sol_quantum}) in terms of Fox $H$-functions
\begin{eqnarray}\label{th_FPeq_sol delta}
N(x,y)&&=\frac{y^{-(1-\nu)(2-\mu)}}{2\pi}\int_{-\infty}^{\infty}E_{\mu,1-(1-\nu)(2-\mu)}\left(\pm y^{\mu}|\kappa|^{\alpha}\right)e^{-\imath\kappa
x}\mathrm{d}\kappa\nonumber\\&&
+\frac{y^{\mu-1}}{2\pi}\int_{-\infty}^{\infty}E_{\mu,\mu}\left(\pm y^{\mu}|\kappa|^{\alpha}\right)
e^{-\imath\kappa x}\mathrm{d}\kappa\nonumber\\&& =
\frac{y^{-(1-\nu)(2-\mu)}}{|x|}H_{3,3}^{2,1}\left[\mp\frac{|x|^{\alpha}}{y^{\mu}}\left|\begin{array}{c
l}
     (1,1), (1-(1-\nu)(2-\mu),\mu), (1,\frac{\alpha}{2})\\
    (1,\alpha), (1,1), (1,\frac{\alpha}{2})
  \end{array}\right.\right]\nonumber\\&&
  +
  \frac{y^{\mu-1}}{|x|}H_{3,3}^{2,1}\left[\mp\frac{|x|^{\alpha}}{y^{\mu}}\left|\begin{array}{c
l}
     (1,1), (\mu,\mu), (1,\frac{\alpha}{2})\\
    (1,\alpha), (1,1), (1,\frac{\alpha}{2})
  \end{array}\right.\right],
\end{eqnarray}
where the upper signs in the solution correspond to the case of fractional Riesz-Feller derivative and lower signs to quantum fractional Riesz-Feller derivative. Moreover, for $\alpha=2$, solution (\ref{th_FPeq_sol delta}) in case of quantum fractional Riesz-Feller derivative becomes
\begin{eqnarray}\label{th_FPeq_sol delta alpha2}
N(x,y)&&=
\frac{y^{-(1-\nu)(2-\mu)}}{2|x|}H_{2,2}^{2,0}\left[\frac{|x|}{y^{\mu/2}}\left|\begin{array}{c
l}
     (1-(1-\nu)(2-\mu),\frac{\mu}{2}), (1,\frac{1}{2})\\
    (1,1), (1,\frac{1}{2})
  \end{array}\right.\right]
  \nonumber\\&&+
  \frac{y^{\mu-1}}{2|x|}H_{2,2}^{2,0}\left[\frac{|x|}{y^{\mu/2}}\left|\begin{array}{c
l}
     (\mu,\frac{\mu}{2}), (1,\frac{1}{2})\\
    (1,1), (1,\frac{1}{2})
  \end{array}\right.\right]\nonumber\\&&
  =\frac{y^{-(1-\nu)(2-\mu)}}{2|x|}H_{1,1}^{1,0}\left[\frac{|x|}{y^{\mu/2}}\left|\begin{array}{c
l}
     (1-(1-\nu)(2-\mu),\frac{\mu}{2})\\
    (1,1)
  \end{array}\right.\right]
 \nonumber\\&&+
  \frac{y^{\mu-1}}{2|x|}H_{1,1}^{1,0}\left[\frac{|x|}{y^{\mu/2}}\left|\begin{array}{c
l}
     (\mu,\frac{\mu}{2})\\
    (1,1)
  \end{array}\right.\right].
\end{eqnarray}
\end{example}

\begin{remark}\label{remark asymptotic FPeq}
For the asymptotic behavior of solution (\ref{th_FPeq_sol delta
alpha2}) for $\frac{|x|}{y^{\mu/2}}\gg1$, we obtain
\begin{eqnarray}\label{asymptotic FPeq}
N(x,y)&&\simeq\frac{\left(\frac{\mu}{2}\right)^{1-2\nu+\frac{1}{2-\mu}}}{2\sqrt{(2-\mu)\pi}}\frac{y^{-(1-\nu)(2-\mu)}}{|x|}\left(\frac{|x|}{y^{\mu/2}}\right)^{\frac{1+2(1-\nu)(2-\mu)}{2-\mu}}
\nonumber\\&&\times\exp\left[-\frac{2-\mu}{2}\left(\frac{\mu}{2}\right)^{\frac{\mu}{2-\mu}}\left(\frac{|x|}{y^{\mu/2}}\right)^{\frac{2}{2-\mu}}\right]\nonumber\\&&
+\frac{\left(\frac{\mu}{2}\right)^{\frac{1-2\mu}{2-\mu}}}{2\sqrt{(2-\mu)\pi}}\frac{y^{\mu-1}}{|x|}\left(\frac{|x|}{y^{\mu/2}}\right)^{\frac{1-2\mu}{2-\mu}}
\exp\left[-\frac{2-\mu}{2}\left(\frac{\mu}{2}\right)^{\frac{\mu}{2-\mu}}\left(\frac{|x|}{y^{\mu/2}}\right)^{\frac{2}{2-\mu}}\right],
\end{eqnarray}
where we employ relations (\ref{H_asymptotic}), (\ref{alpha}),
(\ref{m}), (\ref{C}) and (\ref{B}).
\end{remark}

\begin{remark}\label{remark series representation FLeq}
From the series representation (\ref{H_expansion}) of Fox
$H$-function, we obtain the following series representation of
solution (\ref{th_FPeq_sol delta alpha2})
\begin{eqnarray}\label{series representation PEq}
N(x,y)&&=\frac{y^{-(1-\nu)(2-\mu)-\frac{\mu}{2}}}{2}\sum_{j=0}^{\infty}\frac{(-1)^{j}}{j!\Gamma\left(1-(1-\nu)(2-\mu)-\frac{\mu}{2}(j+1)\right)}\left(\frac{|x|}{y^{\mu/2}}\right)^{j}
\nonumber\\&&+\frac{y^{\frac{\mu}{2}-1}}{2}\sum_{j=0}^{\infty}\frac{(-1)^{j}}{j!\Gamma\left(-\mu
j\right)}\left(\frac{|x|}{y^{\mu/2}}\right)^{j}\nonumber\\&&=
\frac{y^{-(1-\nu)(2-\mu)-\frac{\mu}{2}}}{2}\phi\left(-\frac{\mu}{2},1-(1-\nu)(2-\mu)-\frac{\mu}{2};-\frac{|x|}{y^{\mu/2}}\right)
\nonumber\\&&+\frac{y^{\frac{\mu}{2}-1}}{2}\phi\left(-\mu,0;-\frac{|x|}{y^{\mu/2}}\right),
\end{eqnarray}
from where by using the first few terms of the series (\ref{series
representation PEq}) we can obtain the asymptotic behavior for
$\frac{|x|}{y^{\mu/2}}\ll1$. Here, $\phi(a,b;z)$ is the Wright
function (\ref{Wright}).
\end{remark}

\begin{example}
For $\Phi(x,y)=\delta(x)\delta(y)$ and boundary conditions $f(x)=0$,
$g(x)=\delta(x)$, for $\theta=0$, from (\ref{HML}) and (\ref{cosine
H}), we obtain the solutions (\ref{th_FPeq_sol}) and (\ref{th_FPeq_sol_quantum}) in the following
form
\begin{eqnarray}\label{th_FPeq_sol delta2}
N(x,y)&&=\frac{y^{1-(1-\nu)(2-\mu)}}{2\pi}\int_{-\infty}^{\infty}E_{\mu,2-(1-\nu)(2-\mu)}\left(\pm y^{\mu}|\kappa|^{\alpha}\right)e^{-\imath\kappa
x}\mathrm{d}\kappa
\nonumber\\&&+\frac{y^{\mu-1}}{2\pi}\int_{-\infty}^{\infty}E_{\mu,\mu}\left(\pm y^{\mu}|\kappa|^{\alpha}\right)
e^{-\imath\kappa x}\mathrm{d}\kappa\nonumber\\&& =
\frac{y^{1-(1-\nu)(2-\mu)}}{|x|}H_{3,3}^{2,1}\left[\mp\frac{|x|^{\alpha}}{y^{\mu}}\left|\begin{array}{c
l}
     (1,1), (2-(1-\nu)(2-\mu),\mu), (1,\frac{\alpha}{2})\\
    (1,\alpha), (1,1), (1,\frac{\alpha}{2})
  \end{array}\right.\right]
  \nonumber\\&&+
  \frac{y^{\mu-1}}{|x|}H_{3,3}^{2,1}\left[\mp\frac{|x|^{\alpha}}{y^{\mu}}\left|\begin{array}{c
l}
     (1,1), (\mu,\mu), (1,\frac{\alpha}{2})\\
    (1,\alpha), (1,1), (1,\frac{\alpha}{2})
  \end{array}\right.\right],
\end{eqnarray}
where the upper signs in the solution correspond to the case of fractional Riesz-Feller derivative and lower signs to quantum fractional Riesz-Feller derivative. For $\alpha=2$ solution (\ref{th_FPeq_sol delta2}) in case of quantum fractional Riesz-Feller derivative becomes
\begin{eqnarray}\label{th_FPeq_sol delta2_alpha2}
N(x,y)=\frac{y^{1-(1-\nu)(2-\mu)}}{2|x|}H_{1,1}^{1,0}\left[\frac{|x|}{y^{\mu/2}}\left|\begin{array}{c
l}
     (2-(1-\nu)(2-\mu),\frac{\mu}{2})\\
    (1,1)
  \end{array}\right.\right]
 +
  \frac{y^{\mu-1}}{2|x|}H_{1,1}^{1,0}\left[\frac{|x|}{y^{\mu/2}}\left|\begin{array}{c
l}
     (\mu,\frac{\mu}{2})\\
    (1,1)
  \end{array}\right.\right].
\end{eqnarray}
The asymptotic behavior and series representation of this solution can be found
in a same way as it was done in Remark \ref{remark asymptotic FLeq} and
Remark \ref{remark series representation FLeq}.
\end{example}

\begin{corollary}\label{FLeq}
The solution of the following fractional form of the Laplace
equation
\begin{equation}\label{th_FLeq}
{_{x}}D_{\theta}^{\alpha}N(x,y)+{_{y}}D_{0+}^{\mu,\nu}N(x,y)=0,
\end{equation}
where $x\in R$, $y\in R^{+}$, $1<\alpha\leq2$,
$|\theta|\leq\min\{\alpha,2-\alpha\}$, $1<\mu\leq2$,
$0\leq\nu\leq1$, with boundary conditions
\begin{subequations}
\begin{equation}\label{boundary1 FLeq}
\left({_{y}}I_{0+}^{(1-\nu)(2-\mu)}N\right)(x,0+)=f(x), \quad
\left(\frac{\mathrm{d}}{\mathrm{d}y}\left({_{y}}I_{0+}^{(1-\nu)(2-\mu)}N\right)\right)(x,0+)=g(x),
\end{equation}
\begin{equation}\label{boundary2 FLeq}
\lim_{x\rightarrow\pm\infty}N(x,y)=0,
\end{equation}
\end{subequations}
is given by
\begin{eqnarray}\label{th_FLeq_sol}
N(x,y)&&=\frac{y^{-(1-\nu)(2-\mu)}}{2\pi}\int_{-\infty}^{\infty}E_{\mu,1-(1-\nu)(2-\mu)}\left(y^{\mu}\psi_{\alpha}^{\theta}(\kappa)\right)\hat{f}(\kappa)e^{-\imath\kappa
x}\mathrm{d}\kappa\nonumber\\&&
+\frac{y^{1-(1-\nu)(2-\mu)}}{2\pi}\int_{-\infty}^{\infty}E_{\mu,2-(1-\nu)(2-\mu)}\left(y^{\mu}\psi_{\alpha}^{\theta}(\kappa)\right)\hat{g}(\kappa)e^{-\imath\kappa
x}\mathrm{d}\kappa,
\end{eqnarray}
where $\hat{f}(\kappa)=\mathcal{F}\left[f(x)\right](\kappa)$,
$\hat{g}(\kappa)=\mathcal{F}\left[g(x)\right](\kappa)$.
\end{corollary}

\textit{Proof.} The proof of Corrolary \ref{FLeq} follows directly from
Theorem \ref{FPeq} if we substitute $\Phi(x,y)=0$.

\begin{remark}\label{FLeq_quantum}
If in equation (\ref{th_FLeq}) instead of fractional Riesz-Feller derivative we use quantum fractional Riesz-Feller derivative we obtain the following equation
\begin{equation}\label{th_FLeq_quantum}
{_{x}}D_{\theta}^{*,\alpha}N(x,y)+{_{y}}D_{0+}^{\mu,\nu}N(x,y)=\Phi(x,y).
\end{equation}
For same boundary conditions as those in Theorem (\ref{FLeq}), the solution of equation (\ref{th_FLeq_quantum}) is given by
\begin{eqnarray}\label{th_FLeq_sol_quantum}
N(x,y)&&=\frac{y^{-(1-\nu)(2-\mu)}}{2\pi}\int_{-\infty}^{\infty}E_{\mu,1-(1-\nu)(2-\mu)}\left(-y^{\mu}\psi_{\alpha}^{\theta}(\kappa)\right)\hat{f}(\kappa)e^{-\imath\kappa
x}\mathrm{d}\kappa\nonumber\\&&
+\frac{y^{1-(1-\nu)(2-\mu)}}{2\pi}\int_{-\infty}^{\infty}E_{\mu,2-(1-\nu)(2-\mu)}\left(-y^{\mu}\psi_{\alpha}^{\theta}(\kappa)\right)\hat{g}(\kappa)e^{-\imath\kappa
x}\mathrm{d}\kappa.
\end{eqnarray}
\end{remark}

\begin{corollary}
Solutions of equation
(\ref{th_FLeq}) and (\ref{th_FLeq_quantum}) in case of Caputo fractional derivative ($\nu=1$),  become
\begin{eqnarray}\label{th_FLeq_sol_cor1}
N(x,y)=\frac{1}{2\pi}\int_{-\infty}^{\infty}E_{\mu}\left(\pm y^{\mu}\psi_{\alpha}^{\theta}(\kappa)\right)\hat{f}(\kappa)e^{-\imath\kappa
x}\mathrm{d}\kappa
+\frac{y}{2\pi}\int_{-\infty}^{\infty}E_{\mu,2}\left(\pm y^{\mu}\psi_{\alpha}^{\theta}(\kappa)\right)\hat{g}(\kappa)e^{-\imath\kappa
x}\mathrm{d}\kappa,
\end{eqnarray}
and for R-L fractional derivative ($\nu=0$)
\begin{eqnarray}\label{th_FLeq_sol_cor2}
N(x,y)&&=\frac{y^{\mu-2}}{2\pi}\int_{-\infty}^{\infty}E_{\mu,\mu-1}\left(\pm y^{\mu}\psi_{\alpha}^{\theta}(\kappa)\right)\hat{f}(\kappa)e^{-\imath\kappa
x}\mathrm{d}\kappa
\nonumber\\&&+\frac{y^{\mu-1}}{2\pi}\int_{-\infty}^{\infty}E_{\mu,\mu}\left(\pm y^{\mu}\psi_{\alpha}^{\theta}(\kappa)\right)\hat{g}(\kappa)e^{-\imath\kappa
x}\mathrm{d}\kappa,
\end{eqnarray}
where the upper signs in the solution correspond to the case of fractional Riesz-Feller derivative and lower signs to quantum fractional Riesz-Feller derivative.
\end{corollary}

\begin{example}
If we use the following boundary conditions $f(x)=\delta(x)$ and
$g(x)=0$, solutions (\ref{th_FLeq_sol}) and (\ref{th_FLeq_sol_quantum}) are given by
\begin{eqnarray}\label{th_FLeq_sol_cor3}
N(x,y)=\frac{y^{-(1-\nu)(2-\mu)}}{2\pi}\int_{-\infty}^{\infty}E_{\mu,1-(1-\nu)(2-\mu)}\left(\pm y^{\mu}\psi_{\alpha}^{\theta}(\kappa)\right)e^{-\imath\kappa
x}\mathrm{d}\kappa,
\end{eqnarray}
which for Caputo fractional derivative become
\begin{eqnarray}\label{th_FLeq_sol_cor3_1}
N(x,y)=\frac{1}{2\pi}\int_{-\infty}^{\infty}E_{\mu}\left(\pm y^{\mu}\psi_{\alpha}^{\theta}(\kappa)\right)e^{-\imath\kappa
x}\mathrm{d}\kappa,
\end{eqnarray}
and for R-L fractional derivative
\begin{eqnarray}\label{th_FLeq_sol_cor3_2}
N(x,y)=\frac{y^{\mu-2}}{2\pi}\int_{-\infty}^{\infty}E_{\mu,\mu-1}\left(\pm y^{\mu}\psi_{\alpha}^{\theta}(\kappa)\right)e^{-\imath\kappa
x}\mathrm{d}\kappa,
\end{eqnarray}
where it is used that $\mathcal{F}\left[\delta(x)\right]=1$, and we use the upper signs in the solution in case of fractional Riesz-Feller derivative and lower signs in case of quantum fractional Riesz-Feller derivative.
\end{example}

\begin{remark}\label{th_FLeq_sol_cor3 H}
From relation between M-L and Fox $H$-function (\ref{HML}), by using
Mellin-cosine transform formula (\ref{cosine H}), for the solution
(\ref{th_FLeq_sol_cor3}) we find
\begin{eqnarray}\label{th_FLeq_sol_cor3 H}
N(x,y)&&=\frac{y^{-(1-\nu)(2-\mu)}}{\pi}\int_{0}^{\infty}\cos(\kappa
x)H_{1,2}^{1,1}\left[\mp y^{\mu}e^{\imath\frac{\theta\pi}{2}}|\kappa|^{\alpha}\left|\begin{array}{c
l}
     (0,1)\\
    (0,1),((1-\nu)(2-\mu),\mu)
  \end{array}\right.\right]\mathrm{d}\kappa\nonumber\\&&
=\frac{y^{-(1-\nu)(2-\mu)}}{|x|}H_{3,3}^{2,1}\left[\mp\frac{|x|^{\alpha}}{y^{\mu}e^{\imath\frac{\theta\pi}{2}}}\left|\begin{array}{c
l}
     (1,1), (1-(1-\nu)(2-\mu),\mu), (1,\frac{\alpha}{2})\\
    (1,\alpha), (1,1), (1,\frac{\alpha}{2})
  \end{array}\right.\right],
\end{eqnarray}
which for $\theta=0$ becomes
\begin{eqnarray}\label{th_FLeq_sol_cor3_H theta0}
N(x,y)&&=\frac{y^{-(1-\nu)(2-\mu)}}{|x|}H_{3,3}^{2,1}\left[\mp\frac{|x|^{\alpha}}{y^{\mu}}\left|\begin{array}{c
l}
     (1,1), (1-(1-\nu)(2-\mu),\mu), (1,\frac{\alpha}{2})\\
    (1,\alpha), (1,1), (1,\frac{\alpha}{2})
  \end{array}\right.\right].
\end{eqnarray}
Note that for $\mu=2$, $\nu=1$ solution (\ref{th_FLeq_sol_cor3_H
theta0}) in case of quantum fractional Riesz-Feller derivative is given by
\begin{eqnarray}\label{th_FLeq_sol_cor3_H theta0_mu2}
N(x,y)=\frac{1}{\alpha|x|}H_{3,3}^{2,1}\left[\frac{|x|}{y^{2/\alpha}}\left|\begin{array}{c
l}
     (1,\frac{1}{\alpha}), (1,\frac{2}{\alpha}), (1,\frac{1}{2})\\
    (1,1), (1,\frac{1}{\alpha}), (1,\frac{1}{2})
  \end{array}\right.\right].
\end{eqnarray}
and for $\alpha=2$ by
\begin{eqnarray}\label{th_FLeq_sol_cor3_H theta0_alpha2}
N(x,y)&&=\frac{y^{-(1-\nu)(2-\mu)}}{2|x|}H_{2,2}^{2,0}\left[\frac{|x|}{y^{\mu/2}}\left|\begin{array}{c
l}
     (1-(1-\nu)(2-\mu),\frac{\mu}{2}), (1,\frac{1}{2})\\
    (1,1), (1,\frac{1}{2})
  \end{array}\right.\right]\nonumber\\&&
  =\frac{y^{-(1-\nu)(2-\mu)}}{2|x|}H_{1,1}^{1,0}\left[\frac{|x|}{y^{\mu/2}}\left|\begin{array}{c
l}
     (1-(1-\nu)(2-\mu),\frac{\mu}{2})\\
    (1,1)
  \end{array}\right.\right],
\end{eqnarray}
where we apply the definition (\ref{H_integral}) and the known
properties of $H$-function \cite{Mathai and Saxena}.
\end{remark}

\begin{remark}\label{remark asymptotic FLeq}
From the results in Remark \ref{remark asymptotic FPeq}, for the
asymptotic behavior of solution (\ref{th_FLeq_sol_cor3_H
theta0_alpha2}) for $\frac{|x|}{y^{\mu/2}}\gg1$, we obtain
\begin{eqnarray}\label{asymptotic LEq}
N(x,y)&&\simeq\frac{\left(\frac{\mu}{2}\right)^{1-2\nu+\frac{1}{2-\mu}}}{2\sqrt{(2-\mu)\pi}}\frac{y^{-(1-\nu)(2-\mu)}}{|x|}\left(\frac{|x|}{y^{\mu/2}}\right)^{\frac{1+2(1-\nu)(2-\mu)}{2-\mu}}
\nonumber\\&&\times\exp\left[-\frac{2-\mu}{2}\left(\frac{\mu}{2}\right)^{\frac{\mu}{2-\mu}}\left(\frac{|x|}{y^{\mu/2}}\right)^{\frac{2}{2-\mu}}\right].
\end{eqnarray}
\end{remark}

\begin{remark}\label{remark series representation FLeq}
The series representation of solution (\ref{th_FLeq_sol_cor3_H
theta0_alpha2}) is given by
\begin{eqnarray}\label{series representation LEq}
N(x,y)&&=\frac{y^{-(1-\nu)(2-\mu)-\frac{\mu}{2}}}{2}\sum_{j=0}^{\infty}\frac{(-1)^{j}}{j!\Gamma\left(1-(1-\nu)(2-\mu)-\frac{\mu}{2}(j+1)\right)}\left(\frac{|x|}{y^{\mu/2}}\right)^{j}
\nonumber\\&&=\frac{y^{-(1-\nu)(2-\mu)-\frac{\mu}{2}}}{2}\phi\left(-\frac{\mu}{2},1-(1-\nu)(2-\mu)-\frac{\mu}{2};-\frac{|x|}{y^{\mu/2}}\right).
\end{eqnarray}
By using the first few terms of series (\ref{series representation
LEq}), we can obtain the asymptotic behavior of solution
(\ref{th_FLeq_sol_cor3_H theta0_alpha2}) for
$\frac{|x|}{y^{\mu/2}}\ll1$.
\end{remark}

\begin{example}
For the following boundary conditions $f(x)=0$ and $g(x)=\delta(x)$,
solutions (\ref{th_FLeq_sol}) and (\ref{th_FLeq_sol_quantum}) are given by
\begin{eqnarray}\label{th_FLeq_sol_g}
N(x,y)&&=\frac{y^{1-(1-\nu)(2-\mu)}}{2\pi}\int_{-\infty}^{\infty}E_{\mu,2-(1-\nu)(2-\mu)}\left(\pm y^{\mu}\psi_{\alpha}^{\theta}(\kappa)\right)e^{-\imath\kappa
x}\mathrm{d}\kappa\nonumber\\&&
=\frac{y^{1-(1-\nu)(2-\mu)}}{|x|}H_{3,3}^{2,1}\left[\mp\frac{|x|^{\alpha}}{y^{\mu}e^{\imath\frac{\theta\pi}{2}}}\left|\begin{array}{c
l}
     (1,1), (2-(1-\nu)(2-\mu),\mu), (1,\frac{\alpha}{2})\\
    (1,\alpha), (1,1), (1,\frac{\alpha}{2})
  \end{array}\right.\right].
\end{eqnarray}
The case with $\theta=0$, $\alpha=2$, and quantum fractional Riesz-Feller derivative, yields
\begin{eqnarray}\label{th_FLeq_sol_g02}
N(x,y)
&&=\frac{y^{1-(1-\nu)(2-\mu)}}{2|x|}H_{2,2}^{2,0}\left[\frac{|x|}{y^{\mu/2}}\left|\begin{array}{c
l}
     (2-(1-\nu)(2-\mu),\frac{\mu}{2}), (1,\frac{1}{2})\\
    (1,1), (1,\frac{1}{2})
  \end{array}\right.\right]\nonumber\\&&
  =\frac{y^{1-(1-\nu)(2-\mu)}}{2|x|}H_{1,1}^{1,0}\left[\frac{|x|}{y^{\mu/2}}\left|\begin{array}{c
l}
     (2-(1-\nu)(2-\mu),\frac{\mu}{2})\\
    (1,1)
  \end{array}\right.\right].
\end{eqnarray}
From solution (\ref{th_FLeq_sol_g02}) in case of Caputo fractional derivative ($\nu=1$) one finds
\begin{eqnarray}\label{th_FLeq_sol_g02 Caputo}
N(x,y)
=\frac{y}{2|x|}H_{2,2}^{2,0}\left[\frac{|x|}{y^{\mu/2}}\left|\begin{array}{c
l}
     (2,\frac{\mu}{2}), (1,\frac{1}{2})\\
    (1,1), (1,\frac{1}{2})
  \end{array}\right.\right]=\frac{y}{2|x|}H_{1,1}^{1,0}\left[\frac{|x|}{y^{\mu/2}}\left|\begin{array}{c
l}
     (2,\frac{\mu}{2})\\
    (1,1)
  \end{array}\right.\right],
\end{eqnarray}
and for R-L fractional derivative ($\nu=0$) \cite{samuel thomas}
\begin{eqnarray}\label{th_FLeq_sol_g02 RL}
N(x,y)
=\frac{y^{\mu-1}}{2|x|}H_{2,2}^{2,0}\left[\frac{|x|}{y^{\mu/2}}\left|\begin{array}{c
l}
     (\mu,\frac{\mu}{2}), (1,\frac{1}{2})\\
    (1,1), (1,\frac{1}{2})
  \end{array}\right.\right]=\frac{y^{\mu-1}}{2|x|}H_{1,1}^{1,0}\left[\frac{|x|}{y^{\mu/2}}\left|\begin{array}{c
l}
     (\mu,\frac{\mu}{2})\\
    (1,1)
  \end{array}\right.\right].
\end{eqnarray}
\end{example}

\begin{remark}\label{fractional wave eq}
Here we note that the considered equation (\ref{th_FPeq_quantum}) can be transformed to the following general space-time fractional wave equation in presence of an external source $\Phi(x,t)$
\begin{equation}\label{fractional wave equation}
{_{t}}D_{0+}^{\mu,\nu}N(x,t)={_{x}}D_{\theta}^{\alpha}N(x,t)+\Phi(x,t),
\end{equation}
where we use fractional Riesz-Feller space derivative ${_{x}}D_{\theta}^{\alpha}=-{_{x}}D_{\theta}^{*,\alpha}$ given by (\ref{feller}), $x\in R$, $t\geq0$, $1<\alpha\leq2$,
$|\theta|\leq\min\{\alpha,2-\alpha\}$, $1<\mu\leq2$,
$0\leq\nu\leq1$, with initial conditions
\begin{subequations}
\begin{equation}\label{initial frac wave eq}
\left({_{t}}I_{0+}^{(1-\nu)(2-\mu)}N\right)(x,0+)=f(x), \quad
\left(\frac{\mathrm{d}}{\mathrm{d}t}\left({_{t}}I_{0+}^{(1-\nu)(2-\mu)}N\right)\right)(x,0+)=g(x),
\end{equation}
and boundary conditions
\begin{equation}\label{boundary frac wave}
\lim_{x\rightarrow\pm\infty}N(x,t)=0.
\end{equation}
\end{subequations}
So, its solution is given by
\begin{eqnarray}\label{frac wave eq_sol}
N(x,t)&&=\frac{t^{-(1-\nu)(2-\mu)}}{2\pi}\int_{-\infty}^{\infty}E_{\mu,1-(1-\nu)(2-\mu)}\left(-t^{\mu}\psi_{\alpha}^{\theta}(\kappa)\right)\hat{f}(\kappa)e^{-\imath\kappa
x}\mathrm{d}\kappa\nonumber\\&&
+\frac{t^{1-(1-\nu)(2-\mu)}}{2\pi}\int_{-\infty}^{\infty}E_{\mu,2-(1-\nu)(2-\mu)}\left(-t^{\mu}\psi_{\alpha}^{\theta}(\kappa)\right)\hat{g}(\kappa)e^{-\imath\kappa
x}\mathrm{d}\kappa\nonumber\\&&
+\frac{1}{2\pi}\int_{-\infty}^{\infty}\left({_{t}}\mathcal{E}_{0+;\mu,\mu}^{-\psi_{\alpha}^{\theta}(\kappa);1}\hat{\Phi}\right)(\kappa,t)
e^{-\imath\kappa x}\mathrm{d}\kappa,
\end{eqnarray}
where
$\hat{\Phi}(\kappa,t)=\mathcal{F}\left[\Phi(x,t)\right](\kappa,t)$. Thus, all the previously obtained results, series representations and asymptotic behaviors in case of quantum fractional Riesz-Feller derivative can be used for this fractional wave equation with fractional Riesz-Feller space derivative and Hilfer-composite fractional time derivative. From this solution many obtained results for fractional wave equations with Caputo or R-L time fractional derivative can be recovered. For example, if $\Phi(x,t)=0$ we obtain the general space-time wave equation ${_{t}}D_{0+}^{\mu,\nu}N(x,t)={_{x}}D_{\theta}^{\alpha}N(x,t)$ which contains a number of limiting cases.
\end{remark}

\section{Fractional Helmholtz equation}

The inhomogeneous Helmholtz equation in two variables is
given by
\begin{equation}\label{inhHelmholtz eq}
\frac{\partial^{2}}{\partial
x^{2}}N(x,y)+\frac{\partial^{2}}{\partial
y^{2}}N(x,y)+k^{2}N(x,y)=\Phi(x,y),
\end{equation}
where $N(x,y)$ is the field variable, $k$ is the wave number, and
$\Phi(x,y)$ is a given function. For $\Phi(x,y)=0$ it becomes
homogeneous Helmholtz equation. Furthermore, if $k=0$ it is related
with the Poisson and Laplace equations considered in previous
section. For a given form of $\Phi(x,y)$, equation
(\ref{inhHelmholtz eq}) corresponds to the time-independent wave
equation, which may be used for modeling vibrating membrane.

Here we analyze fractional generalization of the inhomogeneous
Helmholtz equation (\ref{inhHelmholtz eq}).

\begin{theorem}\label{FHeq}
The solution of the following inhomogeneous fractional Helmholtz
equation
\begin{equation}\label{th_FHeq}
{_{x}}D_{\theta}^{\alpha}N(x,y)+{_{y}}D_{0+}^{\mu,\nu}N(x,y)+k^{2}N(x,y)=\Phi(x,y),
\end{equation}
where $x\in R$, $y\geq0$, $1<\alpha\leq2$,
$|\theta|\leq\min\{\alpha,2-\alpha\}$, $1<\mu\leq2$,
$0\leq\nu\leq1$, with boundary conditions
\begin{subequations}
\begin{equation}\label{boundary1 FHeq}
\left({_{y}}I_{0+}^{(1-\nu)(2-\mu)}N\right)(x,0+)=f(x), \quad
\left(\frac{\mathrm{d}}{\mathrm{d}y}\left({_{y}}I_{0+}^{(1-\nu)(2-\mu)}N\right)\right)(x,0+)=g(x),
\end{equation}
\begin{equation}\label{boundary2 FHeq}
\lim_{x\rightarrow\pm\infty}N(x,y)=0,
\end{equation}
\end{subequations}
is given by
\begin{eqnarray}\label{th_FHeq_sol}
N(x,y)&&=\frac{y^{-(1-\nu)(2-\mu)}}{2\pi}\int_{-\infty}^{\infty}E_{\mu,1-(1-\nu)(2-\mu)}\left(y^{\mu}\left(\psi_{\alpha}^{\theta}(\kappa)-k^{2}\right)\right)\hat{f}(\kappa)e^{-\imath\kappa
x}\mathrm{d}\kappa\nonumber\\&&
+\frac{y^{1-(1-\nu)(2-\mu)}}{2\pi}\int_{-\infty}^{\infty}E_{\mu,2-(1-\nu)(2-\mu)}\left(y^{\mu}\left(\psi_{\alpha}^{\theta}(\kappa)-k^{2}\right)\right)\hat{g}(\kappa)e^{-\imath\kappa
x}\mathrm{d}\kappa\nonumber\\&&
+\frac{1}{2\pi}\int_{-\infty}^{\infty}\left({_{y}}\mathcal{E}_{0+;\mu,\mu}^{\psi_{\alpha}^{\theta}(\kappa)-k^{2};1}\hat{\Phi}\right)(\kappa,y)
e^{-\imath\kappa x}\mathrm{d}\kappa,
\end{eqnarray}
where
$\hat{\Phi}(\kappa,y)=\mathcal{F}\left[\Phi(x,y)\right](\kappa,y)$.
\end{theorem}

\textit{Proof.} In a same way as previously, by Laplace transform
(\ref{generalized hilfer_laplace}) to equation (\ref{th_FHeq}) and
from the boundary conditions (\ref{boundary1 FHeq}), we obtain
\begin{equation}\label{sol3}
{_{x}}D_{\alpha}^{\theta}\tilde{N}(x,s)+s^{\mu}\tilde{N}(x,s)-s^{1-\nu(2-\mu)}f(x)-s^{-\nu(2-\mu)}g(x)+k^{2}\tilde{N}(x,s)=\tilde{\Phi}(x,s),
\end{equation}
where $\tilde{N}(x,s)=\mathcal{L}\left[N(x,t)\right]$. The Fourier
transform (\ref{quantum RF}) to (\ref{sol3}) yields
\begin{equation}\label{sol22}
\hat{\tilde{N}}(\kappa,s)=\frac{s^{1-\nu(2-\mu)}}{s^{\mu}-\left(\psi_{\alpha}^{\theta}(\kappa)-k^{2}\right)}\hat{f}(\kappa)
+\frac{s^{-\nu(2-\mu)}}{s^{\mu}-\left(\psi_{\alpha}^{\theta}(\kappa)-k^{2}\right)}\hat{g}(\kappa)+\frac{1}{s^{\mu}-\left(\psi_{\alpha}^{\theta}(\kappa)-k^{2}\right)}\hat{\tilde{\Phi}}(\kappa,s),
\end{equation}
where
$\hat{\tilde{\Phi}}(\kappa,s)=\mathcal{F}\left[\tilde{\Phi}(x,s)\right]$.
By application of Lemma (\ref{lemma_ML}) and Lemma (\ref{lemma_ML
convolution}), by inverse Fourier transform we obtain solution
(\ref{th_FHeq_sol}).

\begin{remark}\label{FHeq_quantum}
If in equation (\ref{th_FHeq}) instead of fractional Riesz-Feller derivative we use quantum fractional Riesz-Feller derivative we obtain the following equation
\begin{equation}\label{th_FHeq_quantum}
{_{x}}D_{\theta}^{*,\alpha}N(x,y)+{_{y}}D_{0+}^{\mu,\nu}N(x,y)+k^{2}N(x,t)=\Phi(x,y).
\end{equation}
If we use same boundary conditions as those used in Theorem (\ref{FHeq}), we obtain the following solution
\begin{eqnarray}\label{th_FHeq_sol_quantum}
N(x,y)&&=\frac{y^{-(1-\nu)(2-\mu)}}{2\pi}\int_{-\infty}^{\infty}E_{\mu,1-(1-\nu)(2-\mu)}\left(-y^{\mu}\left(\psi_{\alpha}^{\theta}(\kappa)+k^{2}\right)\right)\hat{f}(\kappa)e^{-\imath\kappa
x}\mathrm{d}\kappa\nonumber\\&&
+\frac{y^{1-(1-\nu)(2-\mu)}}{2\pi}\int_{-\infty}^{\infty}E_{\mu,2-(1-\nu)(2-\mu)}\left(-y^{\mu}\left(\psi_{\alpha}^{\theta}(\kappa)+k^{2}\right)\right)\hat{g}(\kappa)e^{-\imath\kappa
x}\mathrm{d}\kappa\nonumber\\&&
+\frac{1}{2\pi}\int_{-\infty}^{\infty}\left({_{y}}\mathcal{E}_{0+;\mu,\mu}^{-\left(\psi_{\alpha}^{\theta}(\kappa)+k^{2}\right);1}\hat{\Phi}\right)(\kappa,y)
e^{-\imath\kappa x}\mathrm{d}\kappa.
\end{eqnarray}
\end{remark}

\begin{corollary}
For $\nu=0$ (R-L fractional derivative) one finds the following
solutions \cite{samuel thomas}
\begin{eqnarray}\label{th_FHeq_sol RL}
N(x,y)&&=\frac{y^{\mu-2}}{2\pi}\int_{-\infty}^{\infty}E_{\mu,\mu-1}\left(y^{\mu}\left(\pm\psi_{\alpha}^{\theta}(\kappa)-k^{2}\right)\right)\hat{f}(\kappa)e^{-\imath\kappa
x}\mathrm{d}\kappa\nonumber\\&&
+\frac{y^{\mu-1}}{2\pi}\int_{-\infty}^{\infty}E_{\mu,\mu}\left(y^{\mu}\left(\pm\psi_{\alpha}^{\theta}(\kappa)-k^{2}\right)\right)\hat{g}(\kappa)e^{-\imath\kappa
x}\mathrm{d}\kappa
\nonumber\\&&+\frac{1}{2\pi}\int_{-\infty}^{\infty}\left({_{y}}\mathcal{E}_{0+;\mu,\mu}^{\pm\psi_{\alpha}^{\theta}(\kappa)-k^{2};1}\hat{\Phi}\right)(\kappa,y)
e^{-\imath\kappa x}\mathrm{d}\kappa,
\end{eqnarray}
and for $\nu=1$ - solutions
\begin{eqnarray}\label{th_FHeq_sol Caputo}
N(x,y)&&=\frac{1}{2\pi}\int_{-\infty}^{\infty}E_{\mu}\left(y^{\mu}\left(\pm\psi_{\alpha}^{\theta}(\kappa)-k^{2}\right)\right)\hat{f}(\kappa)e^{-\imath\kappa
x}\mathrm{d}\kappa\nonumber\\&&
+\frac{y}{2\pi}\int_{-\infty}^{\infty}E_{\mu,2}\left(y^{\mu}\left(\pm\psi_{\alpha}^{\theta}(\kappa)-k^{2}\right)\right)\hat{g}(\kappa)e^{-\imath\kappa
x}\mathrm{d}\kappa
\nonumber\\&&+\frac{1}{2\pi}\int_{-\infty}^{\infty}\left({_{y}}\mathcal{E}_{0+;\mu,\mu}^{\pm\psi_{\alpha}^{\theta}(\kappa)-k^{2};1}\hat{\Phi}\right)(\kappa,y)
e^{-\imath\kappa x}\mathrm{d}\kappa,
\end{eqnarray}
where the upper signs in the solution correspond to the case of fractional Riesz-Feller derivative and lower signs to quantum fractional Riesz-Feller derivative.
\end{corollary}

\begin{corollary}
The solutions of equations (\ref{th_FHeq}) and (\ref{th_FHeq_quantum}) for a source term of form
$\Phi(x,y)=\delta(x)\frac{y^{-\beta}}{\Gamma(1-\beta)}$ are given by
\begin{eqnarray}\label{th_FHeq_sol delta power}
N(x,y)&&=\frac{y^{-(1-\nu)(2-\mu)}}{2\pi}\int_{-\infty}^{\infty}E_{\mu,1-(1-\nu)(2-\mu)}\left(y^{\mu}\left(\pm\psi_{\alpha}^{\theta}(\kappa)-k^{2}\right)\right)\hat{f}(\kappa)e^{-\imath\kappa
x}\mathrm{d}\kappa\nonumber\\&&
+\frac{y^{1-(1-\nu)(2-\mu)}}{2\pi}\int_{-\infty}^{\infty}E_{\mu,2-(1-\nu)(2-\mu)}\left(y^{\mu}\left(\pm\psi_{\alpha}^{\theta}(\kappa)-k^{2}\right)\right)\hat{g}(\kappa)e^{-\imath\kappa
x}\mathrm{d}\kappa\nonumber\\&&
+\frac{y^{\mu-\beta}}{2\pi}\int_{-\infty}^{\infty}E_{\mu,\mu-\beta+1}\left(y^{\mu}\left(\pm\psi_{\alpha}^{\theta}(\kappa)-k^{2}\right)\right)
e^{-\imath\kappa x}\mathrm{d}\kappa,
\end{eqnarray}
where the upper signs in the solution correspond to the case of fractional Riesz-Feller derivative and lower signs to quantum fractional Riesz-Feller derivative.
\end{corollary}

\begin{example}\label{ex FHeq}
The solution of the following fractional Helmholtz equation
\begin{equation}\label{th_FHeq delta}
{_{x}}D_{\theta}^{\alpha}N(x,y)+{_{y}}D_{0+}^{\mu,\nu}N(x,y)+k^{2}N(x,y)=\delta(x)\delta(y),
\end{equation}
where $x\in R$, $x\geq0$, $1<\alpha\leq2$,
$|\theta|\leq\min\{\alpha,2-\alpha\}$, $1<\mu\leq2$,
$0\leq\nu\leq1$, with boundary conditions
\begin{subequations}
\begin{equation}\label{boundary1 FHeq delta}
\left({_{y}}I_{0+}^{(1-\nu)(2-\mu)}N\right)(x,0+)=\delta(x), \quad
\left(\frac{\mathrm{d}}{\mathrm{d}y}\left({_{y}}I_{0+}^{(1-\nu)(2-\mu)}N\right)\right)(x,0+)=0,
\end{equation}
\begin{equation}\label{boundary2 FHeq delta}
\lim_{x\rightarrow\pm\infty}N(x,y)=0,
\end{equation}
\end{subequations}
is given by
\begin{eqnarray}\label{th_FHeq_sol delta}
N(x,y)&&=\frac{y^{-(1-\nu)(2-\mu)}}{2\pi}\int_{-\infty}^{\infty}E_{\mu,1-(1-\nu)(2-\mu)}\left(y^{\mu}\left(\psi_{\alpha}^{\theta}(\kappa)-k^{2}\right)\right)e^{-\imath\kappa
x}\mathrm{d}\kappa\nonumber\\&&
+\frac{y^{\mu-1}}{2\pi}\int_{-\infty}^{\infty}E_{\mu,\mu}\left(y^{\mu}\left(\psi_{\alpha}^{\theta}(\kappa)-k^{2}\right)\right)
e^{-\imath\kappa x}\mathrm{d}\kappa.
\end{eqnarray}
\end{example}

\begin{example}
If the boundary conditions are given by $f(x)=0$ and
$g(x)=\delta(x)$, equation from Example (\ref{ex FHeq}) has a
solution of form
\begin{eqnarray}\label{th_FHeq_sol 0 delta}
N(x,y)&&=\frac{y^{1-(1-\nu)(2-\mu)}}{2\pi}\int_{-\infty}^{\infty}E_{\mu,2-(1-\nu)(2-\mu)}\left(y^{\mu}\left(\psi_{\alpha}^{\theta}(\kappa)-k^{2}\right)\right)e^{-\imath\kappa
x}\mathrm{d}\kappa\nonumber\\&&
+\frac{y^{\mu-1}}{2\pi}\int_{-\infty}^{\infty}E_{\mu,\mu}\left(y^{\mu}\left(\psi_{\alpha}^{\theta}(\kappa)-k^{2}\right)\right)
e^{-\imath\kappa x}\mathrm{d}\kappa.
\end{eqnarray}
\end{example}

\begin{remark}\label{fractional wave eq with additional term}
Note that equation (\ref{th_FHeq_quantum}) can be transformed to the following general space-time fractional wave equation in presence of an external source $\Phi(x,t)$
\begin{equation}\label{fractional wave equation additional term}
{_{t}}D_{0+}^{\mu,\nu}N(x,t)={_{x}}D_{\theta}^{\alpha}N(x,t)-k^{2}N(x,t)+\Phi(x,t),
\end{equation}
where we use fractional Riesz-Feller space derivative ${_{x}}D_{\theta}^{\alpha}=-{_{x}}D_{\theta}^{*,\alpha}$ given by (\ref{feller}), $x\in R$, $t\geq0$, $1<\alpha\leq2$,
$|\theta|\leq\min\{\alpha,2-\alpha\}$, $1<\mu\leq2$,
$0\leq\nu\leq1$, with initial conditions
\begin{subequations}
\begin{equation}\label{initial frac wave eq add}
\left({_{t}}I_{0+}^{(1-\nu)(2-\mu)}N\right)(x,0+)=f(x), \quad
\left(\frac{\mathrm{d}}{\mathrm{d}t}\left({_{t}}I_{0+}^{(1-\nu)(2-\mu)}N\right)\right)(x,0+)=g(x),
\end{equation}
and boundary conditions
\begin{equation}\label{boundary frac wave add}
\lim_{x\rightarrow\pm\infty}N(x,t)=0.
\end{equation}
\end{subequations}
Thus, its solution is given by
\begin{eqnarray}\label{frac wave eq_sol additional term}
N(x,t)&&=\frac{t^{-(1-\nu)(2-\mu)}}{2\pi}\int_{-\infty}^{\infty}E_{\mu,1-(1-\nu)(2-\mu)}\left(-t^{\mu}\left(\psi_{\alpha}^{\theta}(\kappa)+k^{2}\right)\right)\hat{f}(\kappa)e^{-\imath\kappa
x}\mathrm{d}\kappa\nonumber\\&&
+\frac{t^{1-(1-\nu)(2-\mu)}}{2\pi}\int_{-\infty}^{\infty}E_{\mu,2-(1-\nu)(2-\mu)}\left(-t^{\mu}\left(\psi_{\alpha}^{\theta}(\kappa)+k^{2}\right)\right)\hat{g}(\kappa)e^{-\imath\kappa
x}\mathrm{d}\kappa\nonumber\\&&
+\frac{1}{2\pi}\int_{-\infty}^{\infty}\left({_{t}}\mathcal{E}_{0+;\mu,\mu}^{-\left(\psi_{\alpha}^{\theta}(\kappa)+k^{2}\right);1}\hat{\Phi}\right)(\kappa,t)
e^{-\imath\kappa x}\mathrm{d}\kappa,
\end{eqnarray}
where $\hat{\Phi}(\kappa,t)=\mathcal{F}\left[\Phi(x,t)\right](\kappa,t)$. This solution contains a number of limiting cases.
\end{remark}

\section{Conclusions}

We consider fractional generalization of the Laplace equation,
Poisson equation and Helmholtz equations in two variables. Since
there is no dependence on the time variable, the solutions of
these equations can be considered as a steady-state solutions. The
fractional derivatives used in this paper are of Riesz-Feller and
Hilfer-composite form. M-L type functions, Fox $H$-functions, and
the Prabhakar integral operator containing two parameter M-L
function in the kernel are used to express solutions analytically.
Several special cases of these equations are investigated.
Asymptotic behavior of solutions is analyzed, and series
expression of solutions are provided. The general space-time fractional
wave equation in presence of an external force is considered as well.

\section*{Appendix: Mittag-Leffler and Fox $H$-functions}

The standard (one parameter) M-L function, introduced by
Mittag-Leffler, is defined by
\cite{erdelyi,dzherbashyan,33,35,samko,hilfer,Mathai and Saxena}
\begin{equation}\label{Eq. 4}
E_{\alpha}(z)=\sum_{k=0}^{\infty}\frac{z^k}{\Gamma(\alpha k+1)},
\end{equation}
where $(z \in \mathbb{C}; \Re(\alpha)>0)$. Later, two parameter M-L
function which was introduced by Wiman, and further analyzed by
Agarwal and Humbert, is given by
\cite{erdelyi,dzherbashyan,33,35,samko,hilfer,Mathai and Saxena}
\begin{equation}\label{Eq.5}
E_{\alpha,\beta}(z)=\sum_{k=0}^{\infty}\frac{z^k}{\Gamma(\alpha
k+\beta)},
\end{equation}
where $(z, \beta \in \mathbb{C}; \Re(\alpha)>0)$. The M-L
functions (\ref{Eq. 4}) and (\ref{Eq.5}) are entire functions of
order $\rho=1/\Re(\alpha)$ and type 1. Note that
$E_{\alpha,1}(z)=E_{\alpha}(z)$. These functions are
generalization of the exponential, hyperbolic and trigonometric
functions since $E_{1,1}(z)=e^z$, $E_{2,1}(z^2)=\cosh(z)$,
$E_{2,1}(-z^2)=\cos(z)$ and $E_{2,2}(-z^2)=\sin(z)/z$.

The Laplace transform of the M-L function is given by
\cite{33,35,samko,hilfer,Mathai and Saxena}
\begin{equation}\label{Eq.6}
\mathcal{L}[t^{\beta-1}E_{\alpha,\beta}(\pm a t^\alpha
)]=\int_{0}^{\infty}e^{-st}t^{\beta-1}E_{\alpha,\beta}(\pm
at^\alpha)\textrm{d}t=\frac{s^{\alpha-\beta}}{s^\alpha\mp a},
\end{equation}
where $ \Re(s)>|a|^{1/\alpha}$.

Prabhakar \cite{prabhakar} introduced the following three parameter
M-L function
\begin{equation}\label{ml3}
E_{\alpha,\beta}^\gamma(z)=\sum_{k=0}^{\infty}\frac{(\gamma)_k}{\Gamma(\alpha
k+\beta)}\frac{z^k}{k!},
\end{equation}
where $\beta, \gamma, z \in \mathbb{C}$, $\Re(\alpha)>0$,
$(\gamma)_{k}$ is the Pochhammer symbol. It is an entire function
of order $\rho=1/\Re(\alpha)$ and type 1. Note that
$E_{\alpha,\beta}^1(z)=E_{\alpha,\beta}(z)$. Later, in \cite{34}
it is used the following four parameter generalized M-L function
\begin{equation}\label{Eq.12}
E_{\alpha,\beta}^{\gamma,\kappa}(z)=\sum_{n=0}^{\infty}\frac{(\gamma)_{\kappa
n}}{\Gamma(\alpha n+\beta)}\cdot \frac{z^n}{n!},
\end{equation}
where $(z, \beta, \gamma \in \mathbb{C}; \Re(\alpha)> \max
\{0,\Re(\kappa)-1\}; \Re(\kappa)>0)$ and $(\gamma)_{\kappa n}$ is a
notation of the Pochhammer symbol, as a kernel of a generalized
integral operator. It is an entire function of order
$\rho=\frac{1}{\Re(\alpha-\kappa)+1}$ and type
$\sigma=\frac{1}{\rho}\left(\frac{\{\Re(\alpha)\}^{\Re(\kappa)}}{\{\Re(\alpha)\}^{\Re(\alpha)}}\right)^\rho$
\cite{34}. Note that
$E_{\alpha,\beta}^{\gamma,1}(z)=E_{\alpha,\beta}^{\gamma}(z)$.

The Fox $H$-function (or simply $H$-function) is defined by the
following Mellin-Barnes integral \cite{Mathai and Saxena}
\begin{eqnarray}\label{H_integral}
H_{p,q}^{m,n}(z)=H_{p,q}^{m,n}\left[z\left|\begin{array}{c l}
    (a_1,A_1),...,(a_p,A_p)\\
    (b_1,B_1),...,(b_q,B_q)
  \end{array}\right.\right]=H_{p,q}^{m,n}\left[z\left|\begin{array}{c l}
    (a_p,A_p)\\
    (b_q,B_q)
  \end{array}\right.\right]=\frac{1}{2\pi\imath}\int_{\Omega}\theta(s)z^{s}\textrm{d}s,\nonumber\\
\end{eqnarray}
where
$\theta(s)=\frac{\prod_{j=1}^{m}\Gamma(b_j-B_js)\prod_{j=1}^{n}\Gamma(1-a_j+A_js)}{\prod_{j=m+1}^{q}\Gamma(1-b_j+B_js)\prod_{j=n+1}^{p}\Gamma(a_j-A_js)}$,
$0\leq n\leq p$, $1\leq m\leq q$, $a_i,b_j \in \mathbb{C}$, $A_i,B_j
\in \mathbb{R}^{+}$, $i=1,...,p$, $j=1,...,q$. The contour $\Omega$
starting at $c-\imath\infty$ and ending at $c+\imath\infty$
separates the poles of the function $\Gamma(b_j+B_js)$, $j=1,...,m$
from those of the function $\Gamma(1-a_i-A_is)$, $i=1,...,n$. The
expansion for the $H$-function (\ref{H_integral}) is given by
\cite{Mathai and Saxena}
\begin{eqnarray}\label{H_expansion}
&&H_{p,q}^{m,n}\left[z\left|\begin{array}{l}
    (a_1,A_1),...,(a_p,A_p)\\
    (b_1,B_1),...,(b_q,B_q)
  \end{array}\right.\right]=\nonumber\\&&
=\sum_{h=1}^{m}\sum_{k=0}^{\infty}\frac{\prod_{j=1, j\neq
h}^{m}\Gamma\left(b_j-B_j\frac{b_h+k}{B_h}\right)\prod_{j=1}^{n}\Gamma\left(1-a_j+A_j\frac{b_h+k}{B_h}\right)}{\prod_{j=m+1}^{q}\Gamma\left(1-b_j+B_j\frac{b_h+k}{B_h}\right)\prod_{j=n+1}^{p}\Gamma\left(a_j-A_j\frac{b_h+k}{B_h}\right)}\cdot\frac{(-1)^kz^{(b_h+k)/B_h}}{k!B_h}.\nonumber\\
\end{eqnarray}

From the Mellin-Barnes integral representation of two parameter
M-L function, one can find the following relation with the Fox
$H$-function \cite{Mathai and Saxena}
\begin{eqnarray}\label{HML}
E_{\alpha,\beta}(z)=\frac{1}{2\pi\imath}\int_{\Omega}\frac{\Gamma(s)\Gamma(1-s)}{\Gamma(\beta-\alpha
s)}z^{s}\textrm{d}s=H_{1,2}^{1,1}\left[-z\left|\begin{array}{c l}
     (0,1)\\
    (0,1),(1-\beta,\alpha)
  \end{array}\right.\right],
\end{eqnarray}
where the contour $\Omega$ starts at $c-\imath\infty$, ends at
$c+\imath\infty$, and separates the poles of functions $\Gamma(s)$
and $\Gamma(1-\beta+\alpha s)$, from those of the function
$\Gamma(1-s)$. It is shown by Mathai, Saxena and Haubold that the
integral converges all $z$ \cite{Mathai and Saxena}.

The Mellin-cosine transform of the $H$-function is given by
\cite{Mathai and Saxena}
\begin{eqnarray}\label{cosine H}
&&\int_{0}^{\infty}k^{\rho-1}\cos(kx)H_{p,q}^{m,n}\left[ak^{\delta}\left|\begin{array}{c
l}
    (a_p,A_p)\\
    (b_q,B_q)
  \end{array}\right.\right]\mathrm{d}k=\nonumber\\&&=\frac{\pi}{x^\rho}H_{q+1,p+2}^{n+1,m}\left[\frac{x^\delta}{a}\left|\begin{array}{c
l}
     (1-b_q,B_q), (\frac{1+\rho}{2}, \frac{\delta}{2})\\
    (\rho,\delta), (1-a_p,A_p), (\frac{1+\rho}{2},\frac{\delta}{2})
  \end{array}\right.\right],
\end{eqnarray}
where $\Re\left(\rho+\delta \min_{1\leq j\leq
m}\left(\frac{b_j}{B_j}\right)\right)>1$, $x^\delta>0$,
$\Re\left(\rho+\delta \max_{1\leq j\leq
n}\left(\frac{a_j-1}{A_j}\right)\right)<\frac{3}{2}$,
$|\arg(a)|<\pi\theta/2$, $\theta>0$,
$\theta=\sum_{j=1}^{n}A_j-\sum_{j=n+1}^{p}A_j+\sum_{j=1}^{m}B_j-\sum_{j=m+1}^{q}B_j$.

The asymptotic expansion of the Fox $H$-function $H_{p,q}^{m,0}(z)$
where $q=m$ for large $z$ is \cite{braaksma,Mathai and Saxena}
\begin{eqnarray}\label{H_asymptotic}
&H_{p,q}^{m,0}(z)\sim
Bz^{(1-\alpha)/m^{\ast}}\exp\left(-m^{\ast}C^{1/m^{\ast}}z^{1/m^{\ast}}\right),
\end{eqnarray}
where
\begin{eqnarray}\label{alpha}
\alpha=\sum_{k=1}^{p}a_{k}-\sum_{k=1}^{q}b_{k}+\frac{1}{2}(q-p+1),
\end{eqnarray}
\begin{eqnarray}\label{m}
m^{\ast}=\sum_{j=1}^{q}B_{j}-\sum_{j=1}^{p}A_{j}>0,
\end{eqnarray}
\begin{eqnarray}\label{C}
C=\prod_{k=1}^{p}A_{k}^{A_{k}}\prod_{k=1}^{q}B_{k}^{-B_{k}},
\end{eqnarray}
\begin{eqnarray}\label{B}
B=(2\pi)^{(m-p-1)/2}C^{(1-\alpha)/m^{\ast}}{m^{\ast}}^{-1/2}\prod_{k=1}^{p}A_{k}^{1/2-a_{k}}\prod_{k=1}^{m}B_{k}^{b_{k}-1/2}.
\end{eqnarray}

Closely related to the Fox $H$-function is the Fox-Wright function
defined by \cite{Mathai and Saxena}
\begin{eqnarray}\label{Fox-Wright}
{_{p}}\Psi_{q}(z)={_{p}}\Psi_{q}\left[z\left|\begin{array}{l}
    (a_1,A_1),...,(a_p,A_p)\\
    (b_1,B_1),...,(b_q,B_q)
  \end{array}\right.\right]
=\sum_{n=0}^{\infty}\frac{\prod_{j=1}^{p}\Gamma\left(a_j+nA_j\right)}{\prod_{j=1}^{q}\Gamma\left(b_j+nB_j\right)}\cdot\frac{z^{n}}{n!},
\end{eqnarray}
which as a special case gives the Wright function \cite{Mathai and
Saxena}
\begin{eqnarray}\label{Wright}
\phi(a,b;z)={_{0}}\Psi_{1}(z)={_{0}}\Psi_{1}\left[z\left|\begin{array}{l}
    \\
    (b,a)
  \end{array}\right.\right]
=\sum_{n=0}^{\infty}\frac{1}{\Gamma\left(b+na\right)}\cdot\frac{z^{n}}{n!}.
\end{eqnarray}


\begin{thebibliography}{00}

\bibitem{braaksma}
B.L.J. Braaksma, {\it Asymptotic expansions and analytic
continuations for a class of Barnes-integrals}, Compos. Math. 15
(1964), pp. 239-341.

\bibitem{caputo} M. Caputo, Elasticita Dissipacione (Bologna: Zanichelli, 1969).

\bibitem{compte}
A. Compte, {\it Stochastic foundations of fractional dynamics},
Phys. Rev. E 53 (1996), pp. 4191-4193.

\bibitem{dzherbashyan}
M.M. Dzherbashyan, Harmonic Analysis and Boundary Value Problems in
the Complex Domain, Vol 65 ed I Gohberg (Basel: Birkhauser, 1993).

\bibitem{erdelyi}
A. Erd\'{e}lyi, W. Magnus, F. Oberhettinger and F.G. Tricomi, Higher
Transcedential Functions {\bf3} (New York, Toronto and London:
McGraw-Hill Book Company, 1955).

\bibitem{feller}
W. Feller, {\it An Introduction to Probability Theory and Its
Applications, Vol. II} (Wiley, New York) 1968.

\bibitem{furati}
K.M. Furati, M.D. Kassim and N.e.-Tatar, {\it Existence and
uniqueness for a problem involving Hilfer fractional derivative},
Comp. Math. Appl. 64 (2012), pp. 1616-1626.

\bibitem{garg}
M. Garg, A. Sharma and P. Manohar, {\it Linear space-time
fractional reaction-diffusion equation with composite fractional
derivative in time}, J. Fract. Calc. Appl. 5 (2014), pp. 114-121.

\bibitem{hmsrivastava1}
Y.-J. Hao, H.M. Srivastava, H. Jafari, and X.-J. Yang, {\it Helmholtz
and Diffusion Equations Associated with Local Fractional Derivative
Operators Involving the Cantorian and Cantor-Type Cylindrical
Coordinates}, Adv. Math. Phys. Article ID 754248 (2013), 2013.

\bibitem{hilfer csf}
R. Hilfer, {\it Fractional dynamics, irreversibility and
ergodicity breaking}, Chaos, Solitons and Fractals 5 (1995), pp.
1475-1484.

\bibitem{hilfer}
R. Hilfer, Application of Fractional Calculus in Physics (Singapore:
World Scientiffic Publishing Company, 2000).

\bibitem{hilfer2}
R. Hilfer, {\it Experimental evidence for fractional time
evolution in glass forming materials}, Chem. Phys. 284 (2002), pp.
399-408.

\bibitem{hilfer fractals}
R. Hilfer, {\it On fractional relaxation}, Fractals 11 (2003), pp.
251-257.

\bibitem{hilfer_LT}
R. Hilfer, Y. Luchko and \v{Z}. Tomovski, {\it Operational method
for the solution of fractional differential equations with
generalized Riemann-Liouville fractional derivatives}, Fract.
Calc. Appl. Anal. 12 (2009), pp. 299-318.

\bibitem{tatar}
S. Al-Homidan, R.A. Ghanam and N.-e. Tatar, {\it On a generalized
diffusion equation arising in petroleum engineering}, Advances in
Difference Equations 2013 (2013), 349.

\bibitem{jumarie}
G. Jumarie, {\it Non-standard analysis and Liouville-Riemann
derivative}, Chaos, Solitons and Fractals 12 (2001), pp.
2577-2587.

\bibitem{kilbas}
A.A. Kilbas, M. Saigo and R.K. Saxena, {\it Generalized
Mittag-Leffler function and generalized fractional calculus
operators}, Integral Transform. Spec. Func. 15 (2004), pp. 31-49.

\bibitem{35}
A.A. Kilbas, H.M. Srivastava and J.J. Trujillo, Theory and
Applications of Fractional Differential Equations, North-Holland
Mathematical Studies, Vol. 204, (Amsterdam: Elsevier, 2006).

\bibitem{kim fcaa}
M.-H. Kim, G.-C. Ri and Hyong-Chol O, {\it Operational method for
solving multi-term fractional differential equations with the
generalized fractional derivatives}, Fract. Calc. Appl. Anal. 17
(2013), pp. 79-95.

\bibitem{kim2}
M.-H. Kim, G.-C. Ri and Hyong-Chol O, {\it Existence and
solution-representation of IVP for LFDE with generalized
Riemann-Liouville fractional derivatives and $n$ terms},
arXiv:1302.2873v5.

\bibitem{liang1}
J. Liang, W. Zhang, Y.Q. Chen and I. Podlubny, Robustness of
boundary control of fractional wave equations with delayed boundary
measurement using fractional order controller and the Smith
predictor, in: {\it Proceedings of 2005 ASME Design Engineering
Technical Conferences}, Long Beach, California, USA, 2005.

\bibitem{liang2}
J. Liang and Y.Q. Chen, {\it Hybrid symbolic and numerical
simulation studies of time-fractional order wave-diffusion
systems}, Int. J. Control 79 (2006), pp. 1462-1470.

\bibitem{luchko jmp}
Y. Luchko, {\it Fractional Schr\"{o}dinger equation for a particle
moving in a potential well}, J. Math. Phys. 54 (2013), 012111.

\bibitem{19}
F. Mainardi, Fractional diffusive waves in viscoelastic solids, in
J.L. Wegner and F.R. Norwood (Editors), Nonlinear Waves in Solids,
Appl. Mech. Rev., Proc. Issue (1994) pp. 93-97.

\bibitem{1}
F. Mainardi, {\it Fractional relaxation-oscillation and fractional
diffusion-wave phenomena}, Chaos, Solitons and Fractals 7 (1996),
pp. 1461-1477.

\bibitem{2}
F. Mainardi, {\it The fundamental solutions for the fractional
diffusion-wave equation}, Appl. Math. Lett. 9 (1996), pp. 23-28.

\bibitem{mainardi}
F. Mainardi, {\it Applications of integral transforms in
fractional diffusion processes}, Integral Transform. Spec. Func.
15 (2004), pp. 477-484.

\bibitem{mainardi jcam}
F. Mainardi and R. Gorenflo, {\it On Mittag-Leffler-type functions
in fractional evolution processes}, J. Comp. Appl. Math. 118
(2000), pp. 283-299.

\bibitem{mainardi-gorenflo}
F. Mainardi and R. Gorenflo, {\it Time-fractional derivatives in
relaxation processes: a tutorial survey}, Fract. Calc. Appl. Anal.
10 (2007), pp. 269-308.

\bibitem{mainardi pagnini saxena}
F. Mainardi, G. Pagnini and R.K. Saxena, {\it Fox $H$ functions in
fractional diffusion}, J. Comput. Appl. Math. 178 (2005), pp.
321-331.

\bibitem{haubold et al}
H.J. Haubold, A.M. Mathai and R.K. Saxena, {\it Solutions of
fractional reaction-diffusion equations in terms of the
$H$-function}, Bull. Astr. Soc. 35 (2007), pp. 681-689.

\bibitem{Mathai and Saxena}
A.M. Mathai, R.K. Saxena and H.J. Haubold, The $H$-function: Theory
and Applications (New York: Springer, 2010).

\bibitem{metzler ffpe}
R. Metzler, E. Barkai, and J. Klafter, {\it Anomalous diffusion
and relaxation close to thermal equilibrium: a fractional
Fokker-Planck equation approach}, Phys. Rev. Lett. 82 (1999), pp.
3563-3567.

\bibitem{metzler1}
R. Metzler and J. Klafter, {\it The random walk's guide to
anomalous diffusion: a fractional dynamics approach}, Phys. Rep.
339 (2000), pp. 1-77.

\bibitem{metzler2}
R. Metzler and J. Klafter, {\it The restaurant at the end of the
random walk: recent developments in the description of anomalous
transport by fractional dynamics}, J. Phys. A: Math. Gen. 37
(2004), pp. R161-R208.

\bibitem{pagnini}
G. Pagnini, A. Mura and F. Mainardi, {\it Generalized fractional
master equation for self-similar stochastic processes modelling
anomalous diffusion}, Int. J. Stoch. Anal. Article ID 427383
(2012), 2012.

\bibitem{33}
I. Podlubny, Fractional Differential Equations, (San Diego: Acad.
Press, 1999).

\bibitem{prabhakar}
T.R. Prabhakar, {\it A singular integral equation with a
generalized Mittag-Leffler function in the kernel}, Yokohama Math.
J. 19 (1971), pp. 7-15.

\bibitem{Rangarajan}
G. Rangarajan and M. Ding, {\it Anomalous diffusion and the first
passage time problem}, Phys. Rev. E 62 (2000), pp. 120-133.

\bibitem{samko}
S.G. Samko, A.A. Kilbas and O.I. Marichev, Fractional Integrals and
Derivatives. Theory and Applications (New York et al: Gordon and
Breach, 1993).

\bibitem{samuel thomas}
M.S. Samuel and A. Thomas, {\it On fractional Helmholtz
equations}, Fract. Calc. Appl. Anal. 13 (2010), pp. 295-308.

\bibitem{sandev jpa2011}
T. Sandev, R. Metzler and \v{Z}. Tomovski, {\it Fractional
diffusion equation with a generalized Riemann-Liouville time
fractional derivative}, J. Phys. A: Math. Theor. 44 (2011), pp.
255203.

\bibitem{sandev jpa2010}
T. Sandev and \v{Z}. Tomovski, {\it The general time fractional
wave equation for a vibrating string}, J. Phys. A: Math. Theor. 43
(2010), pp. 055204.

\bibitem{luchko epj}
B. Al-Saqabi, L. Boyadjiev and Y. Luchko, {\it Comments on
employing the Riesz-Feller derivative in the Schr\"{o}dinger
equation}, Eur. Phys. J. 222 (2013), pp. 1779-1794.

\bibitem{saxena master eq}
R.K. Saxena, {\it On a fractional master equation and a fractional
diffusion equation}, Mathematics and Statistics 1 (2013), pp.
59-63.

\bibitem{saxena1}
R.K. Saxena, A.M. Mathai and H.J. Haubold, {\it Unified fractional
kinetic equation and a fractional diffusion equation}, Astrophys.
Space Sci. 290 (2004), pp. 299-310.

\bibitem{saxena}
R.K. Saxena, A.M. Mathai and H.J. Haubold, {\it Fractional
reaction-diffusion equations}, Astrophys. Space Sci. 305 (2006),
pp. 289-296.

\bibitem{saxena2}
R.K. Saxena, A.M. Mathai and H.J. Haubold, {\it Solutions of
fractional reaction-diffusion equations in terms of Mittag-Leffler
functions}, Int. J. Sci. Research 17 (2008), pp. 1-17.

\bibitem{saxena arxiv}
R.K. Saxena, A.M. Mathai and H.J. Haubold, {\it Computational
solutions of unified fractional reaction-diffusion equations with
composite fractional time derivative}, arXiv:1210.1453v1.

\bibitem{srivastava2}
H.M. Srivastava and R.K. Saxena, {\it Operators of fractional
integration and their applications}, Appl. Math Comput. 118
(2001). pp. 1-52.

\bibitem{34}
H.M. Srivastava and \v{Z}. Tomovski, {\it Fractional calculus with
an integral operator containing a generalized Mittag-Leffler
function in the kernel}, Appl. Math. Comput. 211 (2009), pp.
198-210.

\bibitem{thomas ijde}
A. Thomas, {\it On a fractional master equation}, Int. J. Diff.
Eq. 2011 (2013), Article ID 346298.

\bibitem{tomovski na}
\v{Z}. Tomovski, {\it Generalized Cauchy type problems for
nonlinear fractional differential equations with composite
fractional derivative operator}, Nonlinear Analysis: Theory,
Methods and Applications 75 (2012), pp. 3364-3384.

\bibitem{tomovski itsf}
\v{Z}. Tomovski, R. Hilfer, and H.M. Srivastava, {\it
Fractional and operational calculus with generalized fractional
derivative operators and Mittag-Leffler type functions}, Integral Transform. Spec. Funct. 21 (2010), pp.
797-814.

\bibitem{tomovski cma}
\v{Z}. Tomovski and T. Sandev, {\it Effects of a fractional
friction with power-law memory kernel on string vibrations},
Comput. Math. Appl. 62 (2011), pp. 1554-1561.

\bibitem{tomovski amc}
\v{Z}. Tomovski and T. Sandev, {\it Fractional wave equation with
a frictional memory kernel of Mittag-Leffler type}, Appl. Math.
Comput. 218 (2012), pp. 10022-10031.

\bibitem{tomovski and sandev nd}
\v{Z}. Tomovski and T. Sandev, {\it Exact solutions for fractional
diffusion equation in a bounded domain with different boundary
conditions}, Nonlinear Dynamics 71 (2013), pp. 671-683.

\bibitem{tomovski et al}
\v{Z}. Tomovski, T. Sandev, R. Metzler, and J. Dubbeldam, {\it
Generalized space-time fractional diffusion equation with
composite fractional time derivative}, Physica A 391 (2012), pp.
2527-2542.

\bibitem{tsallis}
C. Tsallis and E.K. Lenzi, {\it Anomalous diffusion: nonlinear
fractional Fokker-Planck equation}, Chem. Phys. 284 (2002), pp.
341-347.



 \end{thebibliography}
\end{document}